\documentclass[a4paper,11pt]{article}
\pdfoutput=1 
\usepackage{jheppub} 
\usepackage{amsfonts,amssymb}
\usepackage{graphics,psboxit,amsmath} 
\usepackage{subfigure}
\usepackage{graphicx}
\usepackage{verbatim}
\usepackage{young}
\usepackage{array,multirow,hhline}
\usepackage{rotating}
\usepackage{nicefrac}
\usepackage{color}
\usepackage{booktabs}
\usepackage{tocbibind}
\usepackage[normalem]{ulem}
\usepackage{datetime}

\def\be{\begin{equation}}
\def\ee{\end{equation}}
\def\ba{\begin{eqnarray}}
\def\ea{\end{eqnarray}}

\def\IR{\relax{\rm I\kern-.18em R}}

\def\inv{^{\raise.0ex\hbox{${\scriptscriptstyle -}$}\kern-.05em 1}}

\title{Holographic perfect fluidity, Cotton energy--momentum  duality and transport properties}

\author[1,2]{Ayan Mukhopadhyay,}
\author[3]{Anastasios C. Petkou,}
\author[1]{P. Marios Petropoulos,}
\author[1]{Valentina Pozzoli,}
\author[4]{Konstadinos Siampos}

\affiliation[1]{
Centre de Physique Th\'eorique,
Ecole Polytechnique, CNRS UMR 7644,\\
Route de Saclay, 91128 Palaiseau Cedex, France}

\affiliation[2]{ Institut de Physique Th\'eorique, 
        CEA, CNRS URA 2306,
        91191 Gif-sur-Yvette, France}

\affiliation[3]{Institute of Theoretical Physics, Department of Physics,
         Aristotle University of Thessaloniki,
         54124 Thessaloniki, Greece}

\affiliation[4]{Service de M\'ecanique et Gravitation, Universit\'e de Mons,
UMONS, 20 Place du Parc,
7000 Mons, Belgium}

\emailAdd{mukhopadhyay,marios,pozzoli@cpht.polytechnique.fr, petkou@physics.auth.gr, konstantinos.siampos@umons.ac.be}
\preprint{CPHT-RR085.1112}

\abstract{
We investigate background metrics for $2+1$-dimensional holographic theories where the equilibrium solution behaves as a perfect fluid, and admits thus a thermodynamic description. We introduce stationary perfect-Cotton geometries, where the Cotton--York tensor takes the form of the energy--momentum tensor of a perfect fluid, \emph{i.e.} they are of Petrov type $\text{D}_\text{t}$. Fluids in equilibrium in such boundary geometries have non-trivial vorticity. The corresponding bulk can be exactly reconstructed to obtain $3+1$-dimensional stationary black-hole solutions with no naked singularities for appropriate values of the black-hole mass. It follows that an infinite number of transport coefficients vanish for holographic fluids.  Our results imply an intimate relationship between black-hole uniqueness and holographic perfect equilibrium. They also point towards a Cotton/energy--momentum tensor duality constraining the fluid vorticity,  as an intriguing boundary manifestation of the bulk mass/nut duality.}

\begin{document}

\maketitle

\newcommand{\eqn}[1]{(\ref{#1})}

\section{Introduction}

The equilibrium is a special state of a system, which is time-independent and where no entropy production takes place. In a generic static or stationary background, the equilibrium state is complicated with presence of temperature and pressure gradients balanced dynamically by energy and momentum exchange between various parts of the system. 

In this paper, we will be interested in \textit{thermodynamic} or \textit{perfect equilibrium,} which is a special equilibrium where all forces generating energy and momentum transfer are absent. Here we can define an appropriate velocity field $u^\mu$ and an appropriate temperature $T$ such that
(\romannumeral1) $T$ is constant and time-independent, (\romannumeral2) 
$u^\mu\partial_\mu$ generates translation along geodesics, and (\romannumeral3)  the energy--momentum tensor takes the perfect-fluid form:
\begin{equation}
\nonumber
T_{\mu\nu} = (\varepsilon + p) u_\mu u_\nu + p \, g_{\mu\nu}, 
\end{equation}
with the energy density $\varepsilon$ and the pressure $p$ being constant, time-independent and determined thermodynamically as functions of $T$.\footnote{It is not hard to see that when all these three requirements are satisfied, the conservation equations $\nabla^\mu T_{\mu\nu} =0$ are also obeyed, meaning no external drive is required to maintain thermodynamic/perfect equilibrium.}
Such an equilibrium can be characterised globally using thermodynamic data alone.

An important question to ask is \textit{in which backgrounds can such thermodynamic/perfect equilibrium be attained}. We will call such backgrounds  as \textit{perfect geometries}.  Investigation of perfect geometries for a given microscopic theory will give a wealth of information on the nature of it's transport properties and correlation functions. The objective of this paper is to do such an investigation for neutral holographic systems. 

The underlying microscopic theories under consideration here have Lorentzian symmetry and live in $2+1$ dimensions. Backgrounds with high degree of symmetry like Minkowski space will be perfect geometries for any such microscopic theory. However, an inhomogeneous stationary background is expected to be a perfect geometry only for special microscopic theories. We will find a large class of inhomogeneous perfect geometries for neutral holographic systems. This will allow us to deduce that a certain class of infinite number of non-dissipative transport coefficients vanish for holographic theories, which is equivalent to finding stringent constraints on the multi-point energy--momentum tensor correlation functions.

On a practical level, we could imagine moving and bending a  $2+1$-dimensional relativistic fluid in such a way that the induced 
metric on it's worldvolume is one of our perfect
geometries. How could we deduce that such a system is amenable to a holographic description, without having much information regarding its microscopic structure? Our suggestion would be to examine -- experimentally if possible -- the system after it equilibrates and see if the temperature is globally constant, and if the stress tensor is that of a perfect
fluid. If so, this would point towards a positive answer to the above question. This is because we would have then verified that an infinite number of transport
coefficients are zero, in agreement with the our prediction from exact black
hole solutions that at equilibrium the fluid experiences no
thermal or mechanical force. On this practical side, we should of course keep in mind that some of the perfect geometries cannot be embedded in $3+1$-dimensional Minkowski space. They still remain interesting at the conceptual level.

Hydrodynamics provides the long wavelength\footnote{This means that the scale on which physical quantities vary is much larger than the mean free path of the microscopic theory, hence a derivative expansion is justifiable.} effective description for many-body quantum systems. In cases where the typical curvature scale of the manifold is larger than the mean free path, the equilibrium can be described as a special solution of the hydrodynamic equations of motion. It is well known that the transport coefficients of hydrodynamics can be classified into dissipative and non-dissipative ones
(see e.g. \cite{Romatschke:2009im, Kovtun:2012rj}). On the one hand, the dissipative ones lead to entropy production, so they play a  r\^ole only in non-equilibrium situations. On the other hand, non-dissipative transport coefficients do not contribute to entropy production, and therefore do play a  r\^ole in determining the equilibrium configuration. Because of the intimate connection between hydrodynamics and the equilibrium, we will be able to constrain the non-dissipative hydrodynamic transport coefficients via our investigation of perfect geometries.

We will further impose a special feature on our perfect geometries endowing them with a generically unique time-like Killing vector of 
unit norm. We will see that the velocity field will align itself with this Killing vector in perfect equilibrium, thus determining it uniquely by geometric data and simplifying our investigation.

Holography asserts the existence of a one-to-one correspondence between states of a conformal fluid theory on the boundary and its dual bulk geometry (see e.g. \cite{Hubeny, Rangamani:2009xk}). The latter is required to be a solution of vacuum Einstein equations with a regular horizon, which in turn implies the existence of a relation between the boundary metric and energy--momentum tensor.  Investigating perfect geometries for holographic systems amounts to fixing the form of the energy--momentum tensor as that of a perfect fluid in equilibrium and looking for the boundary geometry that gives regular bulk solution. Such a procedure is rather unconventional in AdS/CFT, where usually one fixes the boundary metric instead. Our investigation will lead us to construct black-hole-like stationary solutions of Einstein's gravity in four dimensions admitting thermodynamic description.

An important clue to our investigation is provided by conformally self-dual gravitational instantons of four-dimensional Euclidean Einstein's gravity with negative cosmological constant. When mapped into Lorentzian signature,  
the self-duality of the Weyl tensor hints at a certain duality between the Cotton--York tensor and the energy--momentum tensor of the boundary geometry 
\cite{Leigh:2007wf, Mansi:2008br, Mansi:2008bs,deHaro:2008gp,Miskovic:2009bm}. This duality implies that the Cotton--York tensor of the boundary geometry is proportional to the energy--momentum tensor, and we call geometries with such property as \emph{perfect-Cotton geometries}. 
We will prove that perfect-Cotton boundary geometries correspond to resummable exact stationary bulk solutions of Einstein's equations. The reverse statement is not true because there exist exact stationary bulk Einstein spaces with non-perfect-Cotton boundaries (see Ref. \cite{PD} and  App.  \ref{solutions}).
This result  is an  important achievement of the present work.
Its relationship with the gravitational duality is also far reaching, although putting it on firmer grounds requires more work. 

Our analysis enables us to go further by noting that, as a rule, perfect-Cotton geometries possess an extra spacelike Killing vector, generating a spatial isometry. Hence, we are able to write the explicit form of the bulk Einstein solutions, recovering known four-dimensional stationary black-hole metrics such as the AdS--Kerr--Taub--NUT, as well as new solutions. 

Finally, our constraints on the transport coefficients will come from the absence of terms in the equilibrium energy--momentum tensor particularly those related to the fluid vorticity and it's derivatives. These would have spoilt the perfect equilibrium which we will show to be intimately connected with black-hole uniqueness.

Recently approaches based on the equilibrium partition function \cite{Banerjee:2012iz, Jensen:2012jh} and on the existence of an entropy current in hydrodynamics \cite{Bhattacharyya:2012nq} have been used to constrain hydrodynamic transport coefficients. These approaches do not assume any special property of the microscopic theory, thus the constraints on transport coefficients deduced from them hold even if the holographic description does not apply. Our constraints on transport coefficients follow holographically from black-hole dynamics. It is very likely that our constraints are special to holographic systems and do not follow from the methods taken before, though they are entirely consistent with them. We will elaborate on this later.

The organisation of the paper is as follows. In Sec. \ref{hydro}, we briefly review equilibrium as a solution of relativistic fluid mechanics and give  necessary and sufficient condition for perfect equilibrium to exist. In Sec. \ref{PR}, we discuss the stationary geometries in Papapetrou--Randers form and the kinematics of fluids in perfect equilibrium in such geometries. In Sec. \ref{pCg}, we study perfect-Cotton geometries and exhibit the extra spatial isometry that allows to classify them. Fluids in perfect equilibrium in such perfect-Cotton geometries will be studied in Sec. \ref{hol}, followed by the uplift of the corresponding boundary data to exact black-hole solutions. A comment on the rigidity theorem is made at this occasion. In Sec. \ref{trco}, we find that an infinite number of transport coefficients should vanish for perfect-Cotton geometries to be exactly upliftable. In Sec. \ref{outlook}, we conclude with a discussion on possible future directions. In Apps. \ref{vfc} and \ref{Weyl}, we review the basic properties on vector-field congruences and Weyl-covariant traceless transverse tensors in hydrodynamics, respectively. Finally, in App. \ref{solutions} we give some examples of explicit bulk solutions. 

\section{Hydrodynamics and the equilibrium} \label{hydro}

We focus here on the $2+1$-dimensional boundary fluid system, presenting briefly its equilibrium description, and then analysing the special case when the equilibrium is given by a perfect fluid.

\boldmath  
\subsection{Relativistic hydrodynamics on $2+1$-dimensional curved backgrounds}\label{fluidmech}
\unboldmath  

In the hydrodynamic limit the energy--momentum tensor $T^{\mu\nu}$ of a neutral fluid is a function of the local temperature $T(x)$, of the velocity field $u^\mu(x)$, of the background metric $g_{\mu\nu}(x)$ and of their covariant derivatives. It is valid when the scale of variation of $u^\mu$ and $T$ and the curvature scale of $g_{\mu\nu}$ is larger than the mean free path. The hydrodynamic equations are simply given by the covariant conservation of the energy--momentum tensor  
\begin{equation}\label{cons}
\nabla_\mu T^{\mu\nu}=0.
\end{equation}
One way to define the basic thermodynamic variables is within the so-called \emph{Landau frame}, where the non-transverse part of the energy--momentum tensor vanishes when the pressure is zero. This implies that $u^\mu$ is an eigenvector of the energy--momentum tensor with the eigenvalue being the local energy density $\varepsilon(x)$, namely $T^\mu_{\phantom{\mu}\nu}u^\nu = - \varepsilon u^\mu$. If we moreover require the velocity field to be a time-like vector of unit norm, then $\mathrm{u}$ is uniquely defined at each point in space and time. Furthermore, we can use the equation of state for static local equilibrium\footnote{For the global equilibrium case, the internal energy is a function of both $T$ and the angular velocity $\Omega$, (which can be defined if the background metric has a Killing vector corresponding to an angular rotation symmetry). In the case of local equilibrium, $\varepsilon$ is a function of $T$ alone because a dependence on $\Omega$ would not be compatible with the derivative expansion. Indeed, $\Omega$ is first-order in derivatives but $\varepsilon$ is zeroth order. The global energy function $E(T, \Omega)$ can be reproduced by integrating the various components of the equilibrium form of $T_{\mu\nu}$ \cite{Bhattacharyya:2007vs}.}  $\varepsilon=\varepsilon (T)$ to define the temperature $T$. 
Once we have defined a local temperature $T$, we can again use the equation of state to define the pressure $p(x)$. A local entropy density $s(x)$ can be also introduced. Both $p$ and $s$ can be readily obtained from the thermodynamic identities: $\varepsilon + p = Ts$ and $\mathrm{d}\varepsilon = T \mathrm{d}s$. In a conformal $2+1$-dimensional system, $\varepsilon$ and $p$ are proportional to $T^3$ while $s$ is proportional to $T^2$.

Under the assumptions above, the energy--momentum tensor of a neutral hydrodynamic system can be expanded in derivatives of the hydrodynamical variables, namely
\begin{equation}
\label{T00}
T^{\mu\nu}=T^{\mu\nu}_{(0)} +T^{\mu\nu}_{(1)}+T^{\mu\nu}_{(2)}+\cdots\,,
\end{equation}
where the subscript denotes the number of covariant derivatives. Note that the inverse length scale introduced by the derivatives is taken to be large compared to the microscopic mean free path. The zeroth order energy--momentum tensor is the so called perfect-fluid energy--momentum tensor:  
\begin{equation}\label{T0}
T_{(0)}^{\mu\nu}=  \varepsilon u^\mu u^\nu +p\,
\Delta^{\mu\nu},
\end{equation}
where $\Delta^{\mu\nu} = u^\mu u^\nu + g^{\mu\nu}$ is the projector onto the space orthogonal to $\mathrm{u}$. This corresponds to a fluid being locally in static equilibrium. 
The conservation of the perfect-fluid energy--momentum tensor leads to the relativistic Euler equations:
\begin{equation}\label{Euler0}
 \begin{cases}
 \nabla_{\mathrm{u}}\varepsilon+ (\varepsilon+p)\Theta=0,
 \\\nabla_{\perp}p -(\varepsilon+p)\mathrm{a}=0,
\end{cases}
\end{equation}
where $\nabla_{\mathrm{u}}= \mathrm{u}\cdot\nabla$, $\Theta  = \nabla\cdot \mathrm{u}$, $\nabla_{\perp\mu} = \Delta_\mu^{\phantom{\mu}\nu}\nabla_\nu$, and $a^\mu = (\mathrm{u}\cdot\nabla) u^\mu$
(more formulas on kinematics of relativistic fluids are collected in App. \ref{vfc}).

The higher-order corrections to the energy--momentum tensor involve the transport coefficients of the fluid. These are phenomenological parameters that encode the microscopic properties of the underlying system. In the context of field theories, they can be obtained from studying correlation functions of the energy--momentum tensor at finite temperature in the low-frequency and low-momentum regime (see for example \cite{Moore:2010bu}). 

Transport coefficients are of two kinds: dissipative and non-dissipative ones. The former potentially contribute to the entropy production in systems evolving out of global thermodynamic equilibrium.\footnote{Local thermodynamic equilibrium will always be assumed in our discussions as it is required for the hydrodynamic description to make sense.}
The phenomenological discussion of hydrodynamic transport is precisely based on the existence of an entropy current whose covariant divergence describes entropy production and hence must be positive-definite. This puts bounds on the dissipative transport coefficients and imposes relations between non-dissipative transport coefficients order by order in the derivative expansion \cite{Bhattacharyya:2012nq}. 
A complete classification of all transport coefficients is clearly a huge task.

In this work we will be interested only in those transport coefficients which can are relevant  in determining whether perfect equilibrium can exist or not. The energy--momentum tensor in equilibrium has no dissipative terms and is invariant under time reversal --  in short,
T-invariant. In a neutral fluid, the transport coefficients depend on the temperature only, thus we cannot have any T-odd terms in the equilibrium energy--momentum tensor. Therefore, for our purpose it will be sufficient to look for T-even non-dissipative transport coefficients as these alone can play a r\^ole in determining the equilibrium. Of course, some T-even non-dissipative tensors, which can appear in the energy--momentum tensor, could also vanish kinematically in perfect equilibrium and thus the corresponding transport coefficients could play no r\^ole in determining if the perfect equilibrium can exist. One example of such kind is the tensor\footnote{For a second rank tensor $A^{\mu\nu}$ we introduce 
$
\langle A^{\mu\nu} \rangle =\frac{1}{2} \Delta^{\mu\alpha}\Delta^{\nu\beta}(A_{\alpha\beta}+ A_{\beta\alpha})- \frac{1}{2}\Delta^{\mu\nu}\Delta^{\alpha\beta}A_{\alpha\beta}
$.\label{lr}} $\left\langle a_\mu a_\nu \right\rangle$, which will vanish in perfect equilibrium because the acceleration $a_\mu$  vanishes in perfect equilibrium due to lack of temperature gradients (see following subsection). We will not be interested in transport coefficients which appear with such tensors.

A conformal fluid has traceless and Weyl-covariant energy--momentum tensor leading, in $2+1$ dimensions, to the relation $\varepsilon = 2p$. Furthermore, the Landau-frame choice requires transversality. These properties need of course to be valid order by order in the derivative expansion, \emph{i.e.}  for every term appearing in \eqref{T00}. In App. \ref{Weyl} we give details of the construction of such Weyl-covariant traceless and transverse tensors. We will here provide a few illustrative examples.

If we do not require parity invariance, at first order in $2+1$ dimensions, we can have only two such tensors, namely $\sigma^{\mu \nu}$ given in \eqref{def22} (or \eqref{def23}) and $\eta_{\vphantom{\lambda}}^{\rho\lambda(\mu}u_\rho\sigma_\lambda^{\hphantom{\lambda}\nu)}$,
where
$\eta^{\mu\nu\rho}=\nicefrac{\epsilon^{\mu\nu\rho}}{\sqrt{-g}}$  is the covariant fully antisymmetric tensor with $\epsilon^{012}=-1$.
The first-order correction to the energy--momentum tensor thus reads:
\begin{equation}\label{T1}
T_{(1)}^{\mu \nu}=-2\eta \sigma^{\mu \nu}
      -\zeta_{\mathrm{H}}\eta_{\vphantom{\lambda}}^{\rho\lambda(\mu}u_\rho\sigma_\lambda^{\hphantom{\lambda}\nu)}.
\end{equation}
The first term in (\ref{T1}) involves the shear viscosity $\eta$, which is a dissipative transport coefficient. The second is present in systems that break parity and involves the non-dissipative rotational-Hall-viscosity coefficient $\zeta_{\mathrm{H}}$  in $2+1$ dimensions. Notice that the bulk-viscosity term $\zeta \Delta^{\mu \nu}\Theta$ or the anomalous term $\tilde{\zeta}\Delta^{\mu\nu}
\eta^{\alpha\beta\gamma}
u_\alpha\nabla_\beta u_\gamma$ cannot appear in a conformal fluid because it is tracefull, namely for conformal fluids $\zeta =\tilde{\zeta}=0$.  

The next-order terms in \eqref{T00} can be worked out for the fluids at hand. One can easily see that there are no T-even tensors at second order which could be relevant for determining the perfect equilibrium. 
But at third order the T-even tensors non-vanishing at equilibrium, which also do not depend on acceleration, shear and expansion are:  
\begin{equation}\label{T3}
T_{(3)}^{\mu \nu}= \gamma_{(3)1} \langle C^{\mu\nu}\rangle  + \gamma_{(3)2} \langle\mathcal{D}^\mu W^\nu \rangle,
\end{equation}
where $C^{\mu\nu}$ is the Cotton--York tensor and the bracket is defined in footnote \ref{lr}.  The Weyl-covariant derivative $\mathcal{D}_\mu$ can be defined in terms of fluid variables only \cite{Loga} (for details see App. \ref{Weyl}), and $W^\mu$ is given by
\begin{equation}\label{VW}
W^\mu = \eta^{\mu\nu\rho}u_\nu V_\rho, \quad V^\mu = \nabla_{\perp}^\alpha \omega_{\phantom{\alpha}\alpha}^{\mu} + u^\mu \omega_{\alpha\beta}\omega^{\alpha\beta},
\end{equation}
with $\omega_{\alpha\beta}$ being the vorticity defined in \eqref{def24} (or  \eqref{def3}). 
At the fourth order in derivative expansion, there will be non-dissipative transport coefficients corresponding to T-invariant tensors like $\langle V_\mu V_\nu\rangle$, $\langle W_\mu W_\nu \rangle $, etc.

\subsection{Perfect equilibrium}\label{perfeq}

Stationary solutions\footnote{It is admitted that a non-relativistic fluid is stationary when its velocity field is time-independent. This is of course an observer-dependent statement. For relativistic fluids, one could make this more intrinsic saying that the velocity field commutes with a globally defined time-like Killing vector, assuming that the later exists.} of the relativistic equations of motion \eqref{cons}, when they exist, describe a fluid in global thermodynamic equilibrium.\footnote{This should not be confused with a steady state, where we have stationarity due to a balance between external driving forces and internal dissipation. Such situations will not be discussed here.}   
The prototype example of such a situation is the one of an inertial fluid in Minkowski background with globally defined constant temperature, energy density and pressure. In this case, irrespective of whether the fluid itself is viscous, its energy--momentum tensor, evaluated at the solution, takes the zeroth-order (perfect) form (\ref{T0}) because all derivatives of the hydrodynamic variables vanish and Eqs. \eqref{Euler0} are satisfied.

Local thermodynamic equilibrium, in general, does not require the zeroth-order equations \eqref{Euler0} to be satisfied.  It is however relevant to  ask: are there other situations where the hydrodynamic description of a system is also perfect \emph{i.e.} the energy--momentum tensor, in equilibrium, takes the perfect form (\ref{T0}) solving Eqs. \eqref{Euler0}? 
As anticipated in the introduction, we call these special configurations \emph{perfect equilibrium states}, 
where global thermodynamic description applies. One should stress, however, that more general equilibrium states can exist, for which only the full equations \eqref{cons} with \eqref{T00} are satisfied. Owing to the fact that global thermodynamic equilibrium is incompatible with entropy production, all dissipative terms in  \eqref{T00} for these more general equilibrium states necessarily  vanish,\footnote{See e.g. \cite{Bhattacharyya:2012nq} for a recent discussion.} either because the dissipative transport coefficients are zero, or because the corresponding tensors vanish kinematically -- requiring in particular a special relationship between the fluid's velocity and the background geometry.   Clearly, for perfect-equilibrium states,  \emph{all} higher-derivative terms in \eqref{T00} are absent, making their realisation more challenging. In the following we will examine the dynamical requirements of the microscopic theory and kinematical requirements of the background for having a perfect equilibrium.

Consider a hydrodynamic system with a stationary background metric, having a unique time-like, normalised Killing vector $\xi=\xi^\mu\partial_\mu$, namely 
\begin{equation}\label{Killing}
\nabla_{(\mu} \xi_{\nu)}=0, \quad \xi_\mu \xi^\mu=-1. 
\end{equation}
Although not exhaustive, these systems are interesting in view of their intimate connection with holography, as we will see in the following. Congruences defined by $\xi_\mu$ have vanishing acceleration, shear and expansion (see App. \ref{vfc}), but non-zero vorticity $\omega = \frac{1}{2}\mathrm{d}\xi \Leftrightarrow 
\omega_{\mu\nu}=\nabla_{\mu}\xi_{\nu}$. Then, it is easy to show that a special solution of the Euler equations (\ref{Euler0}) is:
\begin{equation}\label{Euler0K}
\mathrm{u}= \xi, \quad T = \,\text{constant}, \quad  \varepsilon = 2p = \, \text{constant}. 
\end{equation}

Let us assume that one perfect-equilibrium solution exists meaning that higher-derivative corrections to the perfect-fluid form of the energy--momentum tensor vanish at equilibrium. We can then show that (\ref{Euler0K}) is the unique perfect-equilibrium solution, if the background has a unique time-like Killing vector field of unit norm. Non-zero shear and expansion can all contribute to dissipation, and should vanish in equilibrium. As noted in the introduction, the vanishing of temperature gradients is crucial for the equilibrium to have a global thermodynamic interpretation. The acceleration therefore vanishes in perfect equilibrium due to lack of temperature gradients. The Killing vector field is the unique vector field with vanishing shear, expansion and acceleration in generic backgrounds under study. Thus the perfect equilibrium is unique in these generic backgrounds.

One important point to note is that in perfect equilibrium we have no frame ambiguity in defining the velocity field. Since the velocity field is geodesic and is aligned with a Killing vector field of unit norm, it describes a unique local frame where all forces (like those induced by a temperature gradient) vanish.

For the above configuration \eqref{Euler0K} to be a perfect-equilibrium state, one must show that all higher-derivative corrections in \eqref{T00} are actually absent. Since the congruence is shearless, the first corrections \eqref{T1} vanish. If higher-order corrections do also vanish, the fluid indeed reaches this specific global equilibrium state, in which it aligns itself with the congruence of the Killing vector field.  For observers whose worldlines are identified with the Killing congruence at hand, the fluid is at rest: the fluid and the observers are comoving. Had the higher-derivative corrections been non-zero, this comoving state with constant temperature would not have been necessarily an equilibrium state as it would not have been a solution of the equations of motion given by (\ref{cons}). Equations \eqref{Euler0} would have been altered, leading in general to $\mathrm{u}=\xi+\delta\mathrm{u}(\mathrm{x})$ and $T=T_0+\delta T(\mathrm{x})$. Such an excursion will be stationary or not depending on whether the non-vanishing corrections to the perfect energy--momentum tensor are non-dissipative or dissipative.

In order to analyse under which conditions on the transport coefficients, perfect-fluid equilibrium \eqref{Euler0K} is realised, we must list, assuming \eqref{Killing}, the Weyl-covariant, traceless and transverse tensors $\mathcal{T}_{\mu\nu}$ that are non-vanishing  and whose divergence is also non-vanishing, 
when evaluated in the perfect-equilibrium solution \eqref{Euler0K}. 

We call such tensors \emph{dangerous tensors}. Their presence can destroy the existence of the perfect-equilibrium solution, unless the corresponding transport coefficients are vanishing. At every order in the derivative expansion we have a finite number of linearly independent dangerous tensors and each one of them is associated with a transport coefficient, which we call \emph{dangerous transport coefficient}. Hence, a necessary and sufficient condition for the existence of perfect equilibrium in backgrounds with a normalised time-like Killing vector field is that \emph{all dangerous transport coefficients vanish}.  The vanishing of the latter is a statement about the underlying microscopic theory about which we can thus gain new non-trivial information.

We will encounter non-trivial special backgrounds (Minkowski space being a trivial example) where no dangerous tensors are present. On the other hand, we will also consider a large class of backgrounds with a unique normalised time-like Killing vector field, which have infinitely many non-zero dangerous tensors; thus we will be able to probe that an infinite number of non-dissipative transport coefficients vanish. Nevertheless, the question of whether our analysis regarding all possible
transport coefficients is exhaustive or not lies beyond the scope of the present work. It is clear that further insight on this matter can only be gained by perturbing the perfect-equilibrium state.

\section{Fluids in Papapetrou--Randers geometries}
\label{PR}

A stationary metric can be written in the generic form
\begin{equation}
\label{Papa}
\mathrm{d}s^2=B^2\left(-(\mathrm{d}t-b_i \mathrm{d}x^i)^2+a_{ij}\mathrm{d}x^i \mathrm{d}x^j\right),
\end{equation}
where $B, b_i, a_{ij}$ are space-dependent but time-independent functions. These metrics were introduced by Papapetrou in \cite{Papapetrou}. 
They will be called hereafter \emph{Papapetrou--Randers}  because they are part of an interesting network of relationships 
involving the Randers form \cite{Randers}, discussed in detail in \cite{Gibbons} and more recently used in \cite{Leigh:2011au,LPP2, NewPaper}. 

In order for the time-like Killing vector $\partial_t$ to be normalised to $-1$, we must restrict ourselves to the case $B=1$.
Then, $\partial_t$ is identified with the generically \emph{unique} normalised time-like Killing vector of the background and draws the geodesic congruence associated with the fluid worldlines. The normalised three-velocity one-form of the stationary perfect fluid is then
\begin{equation}\label{velform}
\mathrm{u}=-\mathrm{d}t+\mathrm{b},
\end{equation}
where $\mathrm{b}=b_i \mathrm{d}x^i$. We will often write the metric \eqref{Papa} as 
\begin{equation}\label{Papas}
\mathrm{d}s^2=-\mathrm{u}^2+\mathrm{d}\ell^2\, , \quad \mathrm{d}\ell^2= a_{ij}\, \mathrm{d}x^i\mathrm{d}x^j.
\end{equation}
We will adopt the convention that hatted quantities will be referring to the two-dimensional positive-definite metric $a_{ij}$, therefore $\hat\nabla$ for the covariant derivative and $\hat R_{ij}\,\mathrm{d}x^i\mathrm{d}x^j=\frac{\hat R}{2}\mathrm{d}\ell^2 $ for the Ricci tensor built out of $a_{ij}$. For later convenience, we introduce the inverse two-dimensional metric $a^{ij}$ and $ b^i$
such that
\begin{equation}
a^{ij}a_{jk}= \delta^i_k,\quad  b^i=a^{ij}b_j.
\end{equation}
The three-dimensional metric components read:
\begin{equation}
g_{00}=-1,\quad g_{0i}= b_i, \quad
g_{ij}=a_{ij}-b_ib_j,
\end{equation}
and those of the inverse metric:
\begin{equation}
g^{00}=a^{ij}b_i b_j-1,\quad g^{0i}=b^i, \quad
g^{ij}=a^{ij}.
\end{equation}
Finally, 
\begin{equation}
\sqrt{-g}=\sqrt{a},
\end{equation}
where $a$ is the determinant of the symmetric matrix with entries $a_{ij}$.

A perfect fluid at equilibrium, or a fluid at perfect equilibrium, whenever this is possible, along the discussion of Sec. (\ref{perfeq}), on a  Papapetrou--Randers background is such that the worldline of every small part of it is aligned with a representative of the congruence tangent to $\partial_t$. Since  $\partial_t$ is a unit-norm Killing vector, the fluid's flow is geodesic, has neither shear, nor expansion, but does have vorticity, which is inherited from the fact that  $\partial_t$ is not hypersurface-orthogonal.\footnote{For this very same reason, Papapetrou--Randers geometries may in general suffer from global hyperbolicity breakdown. This occurs whenever regions exist where $b_i b^i>1$. There, constant-$t$ surfaces cease being space-like, and potentially exhibit closed time-like curves. This issues were discussed in detail in \cite{Leigh:2011au,LPP2, NewPaper}.} Using \eqref{velform} and \eqref{def3} we find that the vorticity can be written as the following two-form
\begin{equation}
\omega = \frac{1}{2}\omega_{\mu\nu}\mathrm{d}x^\mu\wedge\mathrm{d}x^\nu=\frac{1}{2}\mathrm{d}\mathrm{b}.
\end{equation}
The Hodge-dual of $\omega_{\mu\nu}$  is
\begin{equation}
\psi^\mu = \eta^{\mu\nu\rho}\omega_{\nu\rho}
\Leftrightarrow
\omega_{\nu\rho} =-\frac{1}{2}\eta_{\nu\rho \mu}\psi^\mu.
\end{equation}
In $2+1$ dimensions it is aligned with the velocity field:
\begin{equation}
\psi^\mu =q u^\mu,
\end{equation}
where 
\begin{equation}
\label{qgen}
q(x)=-\frac{\epsilon^{ij}\partial_ib_j}{\sqrt{a}}
\end{equation}
is a static scalar field that we call the \emph{vorticity strength}, carrying dimensions of inverse length. Together  with $\hat R(x)$, the above scalar carries all relevant information for the curvature of the Papapetrou--Randers geometry. We quote for latter use the three-dimensional curvature scalar:
\begin{equation}
R = \hat R+\frac{q^2}{2},
\label{Rsgen}
\end{equation}
the three-dimensional Ricci tensor 
\begin{equation}
R_{\mu\nu}\,\mathrm{d}x^\mu\mathrm{d}x^\nu=
 \frac{q^2}{2}\mathrm{u}^2 +\frac{\hat R+q^2}{2}\mathrm{d}\ell^2
-\mathrm{u}\, \mathrm{d}x^\rho u^\sigma \eta_{\rho\sigma\mu} \nabla^\mu q,
\label{Rgen}
\end{equation}
as well as the  three-dimensional Cotton--York  tensor (often called Cotton in short) \cite{Moutsopoulos:2011ez}:
\begin{eqnarray}
C_{\mu\nu}\,\mathrm{d}x^\mu\mathrm{d}x^\nu&=& \frac{1}{2}\left(\hat\nabla^2 q+\frac{q}{2}(\hat R+2q^2)\right)
\left(2\mathrm{u}^2 +\mathrm{d}\ell^2\right)\nonumber
\\ \nonumber
&&-\frac{1}{2}\left(\hat\nabla_i\hat\nabla_j q\,  \mathrm{d}x^i \mathrm{d}x^j  +\hat\nabla^2 q
\, \mathrm{u}^2\right) \\
&&-\frac{\mathrm{u}}{2}\mathrm{d}x^\rho u^\sigma \eta_{\rho\sigma\mu} \nabla^\mu (\hat R +3 q^2).
\label{Cgen}
\end{eqnarray}
The latter is a symmetric and traceless tensor defined in general as
\begin{equation}
\label{PRCot}
C^{\mu \nu}=\eta^{\mu\rho\sigma}
\nabla_\rho\left(R^{\nu}_{\hphantom{\nu}\sigma}-\frac{1}{4}R\delta^{\nu}_\sigma \right).
\end{equation}
In three-dimensional geometries it replaces the always vanishing Weyl tensor. In particular, conformally flat backgrounds have zero Cotton--York tensor
and vice versa.

The fluid  in perfect equilibrium on Papapetrou--Randers backgrounds has the energy--momentum tensor
\begin{equation}
T^{(0)}_{\mu\nu}\,\mathrm{d}x^\mu\mathrm{d}x^\nu=p\left(2\mathrm{u}^2 +\mathrm{d}\ell^2\right),
\label{T0c}
\end{equation}
with the velocity form being given by \eqref{velform} and $p$ constant. We have used here $\varepsilon = 2p$. We recall that $\varepsilon$ has dimensions of energy density or equivalently $(\mathrm{length})^{-3}$, therefore the energy--momentum tensor and the Cotton--York tensor have the same natural dimensions. This is crucial for the following. 

As discussed in the previous section, the fluid can attain perfect equilibrium if
and only if all the dangerous transport coefficients vanish. It is not hard to see that this will imply constraints on transport coefficients as these stationary backgrounds will generically have infinitely many associated dangerous tensors. For example, there exist non-vanishing tensor structures involving $\nabla^n\mathrm{u},  \nabla^{n}q$ with $n>1$,
which are traceless, transverse and Weyl-covariant, with $q$ being evaluated on the perfect-fluid solution as in \eqref{qgen}. One simple example of such a tensor is $\langle \mathcal{D}_\mu W_\nu \rangle$ (one of the terms in \eqref{T3}) which, when evaluated with $\mathrm{u}$ as in \eqref{velform}, will be given in terms of covariant derivatives of $q$. Also generically $\nabla^\mu \langle \mathcal{D}_\mu W_\nu \rangle \neq 0$ in these stationary backgrounds. This is a dangerous tensor and the corresponding dangerous transport coefficient must vanish, in order that the fluid can attain perfect equilibrium.

When the stationary background has additional isometries, most Weyl-covariant, traceless and  transverse tensors built from derivatives of $\mathrm{u}$ as given by \eqref{velform} will vanish. This will make it hard for dangerous tensors to exist. 
In the following Sec. \ref{TMG} and \ref{Gen1}, we will find examples with
 (\romannumeral1) homogeneous and  (\romannumeral2) axisymmetric spaces, where indeed such a conspiracy will happen. Many possible dangerous tensors will vanish, therefore the corresponding transport coefficients need not vanish in order for perfect equilibrium to exist. As expected, the higher the symmetry of a background, the less number of transport coefficients we will be able to probe by demanding that perfect equilibrium should exist.

\section{Perfect-Cotton geometries}\label{pCg}

The existence of perfect geometries is an issue unrelated to holography. However, in the context of the fluid/gravity correspondence a special class of Papapetrou--Randers background geometries for holographic fluids are perfect geometries. In this section we prove that this is the case if the Cotton--York tensor of the boundary metric takes the form 
\begin{equation}\label{perfCott}
C_{\mu\nu}= \frac{c}{2}(3 u_\mu u_\nu + g_{\mu\nu}), 
\end{equation}
where $c$ is a constant with the dimension of an energy density. This form is known in the literature as Petrov class $\text{D}_{\mathrm{t}}$.\footnote{The subscript t stands for \emph{time-like} and refers to the nature of the vector u. For an exhaustive review on Petrov  \& Segre classification of three-dimensional geometries see e.g. \cite{Chow:2009km}.} We call such geometries \emph{perfect-Cotton geometries} because \eqref{perfCott} has the form of a perfect-fluid energy--momentum tensor, and we present here their properties, as well as  their complete classification  based on the fact that in these geometries an extra spatial isometry is always present. Moreover, perfect-Cotton geometries appear as boundaries of $3+1$-dimensional exact Einstein spaces, which will be studied in the next section.

\subsection{Definition}\label{PCPR}

Consider a Papapetrou--Randers metric \eqref{Papa}. Requiring its Cotton--York tensor \eqref{Cgen} to be of the form \eqref{perfCott} is equivalent to impose the conditions: 
\begin{eqnarray}
\label{cotequileq1}
\hat\nabla^2 q+q(\delta-q^2)&=&2c\\\label{cotequileq2}
a_{ij}\left(\hat\nabla^2 q +\frac{q}{2}(\delta-q^2)-c\right)
&=& \hat\nabla_i\hat\nabla_j q\\
\label{Requil}
\hat R +3 q^2&=&\delta, 
\end{eqnarray}
with $\delta$ being a constant relating the curvature of the two-dimensional base space $\hat R$ with the vorticity strength $q$.

 It is remarkable that perfect-Cotton geometries \emph{always} possess an extra space-like Killing vector.\footnote{We would like to thank Jakob Gath for pointing out this property.}
To prove this we rewrite \eqref{cotequileq1} and \eqref{cotequileq2}
as
\begin{equation}
\label{cotequileq3}
\left(\hat\nabla_i\hat \nabla_j-\frac12\,a_{ij} \hat\nabla^2\right)q=0.
\end{equation}
Any two-dimensional metric can be locally written as 
\begin{equation}
\label{2dcomplex}
\mathrm{d}\ell^2=2\mathrm{e}^{2\Omega(z,\bar z)}\mathrm{d}z\,\mathrm{d}\bar z,
\end{equation}
where $z$ and $\bar z$ are complex-conjugate coordinates. Plugging \eqref{2dcomplex} in \eqref{cotequileq3} we find that the non-diagonal 
equations are always satisfied (tracelessness of the Cotton--York tensor), while the diagonal ones read:
\begin{equation}
\label{cotequileq4}
\partial_z^2q=2\partial_z \Omega\partial_z q\, ,\quad \partial_{\bar z}^2q=2\partial_{\bar z} \Omega\partial_{\bar z} q.
\end{equation}
The latter can be integrated to obtain 
\begin{equation}
\label{cotequileq5}
\partial_zq=\mathrm{e}^{2\Omega-2\bar C(\bar z)}\, ,\quad \partial_{\bar z}q=\mathrm{e}^{2\Omega-2C(z)}
\end{equation}
with $C(z)$ an arbitrary holomorphic function and $\bar C(\bar z)$ its complex conjugate. Trading these functions for
\begin{equation}
w(z)=\int \mathrm{e}^{2 C(z)}\mathrm{d}z\,,\quad \bar w(\bar z)=\int \mathrm{e}^{2 \bar C(\bar z)}\mathrm{d}\bar z,
\end{equation}
and introducing new coordinates $(X,Y)$ as
\begin{equation}
X=w(z)+\bar w(\bar z)\,, \quad Y = i\left(\bar w(\bar z)-w(z)\right), 
\end{equation}
we find using  \eqref{cotequileq5} that the vorticity strength depends only on $X$: $q=q(X)$.
Hence, \eqref{2dcomplex} reads: 
\begin{equation}
\text{d}\ell^2=\frac{1}{2}\partial_X q\left(\mathrm{d}X^2+\mathrm{d}Y^2 \right).
\end{equation}
This condition enforces the existence of an extra Killing vector.
Finally we note that \eqref{Requil} can be obtained by differentiating \eqref{cotequileq1} with respect to $X$.

Without loss of generality, we can choose the two-dimensional coordinates $x$ and $y$ in such a way that the base metric $a_{ij}$ is diagonal
\begin{equation}
\label{gen-2D}
\mathrm{d}\ell^2 = A^2 (x,y) \mathrm{d}x^2 + B^2 (x,y) \mathrm{d}y^2
\end{equation}
and that the spatial component of the velocity vector takes the form:
\begin{equation}
\label{gen-gaug}
\mathrm{b}=b(x,y)\,\mathrm{d}y.
\end{equation}
The important thing to keep in mind is that \eqref{gen-2D} has necessarily an isometry despite its explicit dependence on $x$ and $y$.
The vorticity strength \eqref{qgen} reads thus
\begin{equation}
\label{gen-q}
q=-\frac{\partial_x b}{AB}.
\end{equation}
Further gauge fixing is possible and will be made when appropriate.\footnote{For example, since any two-dimensional space is conformally flat it is possible to set $A=B$. We should however stress that all these choices are local, and the range of coordinates should be treated with care in order to avoid e.g. conical singularities.} The explicit form of Eqs. \eqref{cotequileq1}--\eqref{Requil} in terms of $A(x,y), B(x,y)$ and $b(x,y)$ is not very illuminating and we do not present it here.

\subsection{Classification of the Papapetrou--Randers perfect-Cotton geometries} \label{Gen1}

The presence of the space-like isometry actually simplifies the perfect-Cotton conditions for Papapetrou--Randers metrics. Without loss of generality, we take the space-like Killing vector to be $\partial_y$ in \eqref{gen-2D} and we choose a representation such that $A^2= \nicefrac{1}{G(x)}$, $B^2=G(x)$
and $b=b(x)$. The metric takes then the form
\begin{equation}
\label{FH-rot-byG}
\mathrm{d}s^2 = -\left(\mathrm{d}t- b(x)\,\mathrm{d}y\right)^2+ 
\frac{\mathrm{d}x^2}{G(x)}   + G(x) \mathrm{d}y^2,
\end{equation}
and we are able to solve \eqref{cotequileq1}--\eqref{Requil} in full generality. The solution is written in terms of 6 arbitrary parameters $\tilde c_i$:
\begin{eqnarray}
\label{b}
b(x)&=&\tilde c_0
+ \tilde c_1 x+\tilde c_2 x^2, \\
\label{G}
G(x) &=& \tilde c_5 + \tilde c_4 x+ \tilde c_3 x^2+\tilde c_2  x^3 \left( 2\tilde c_1+\tilde c_2 x\right).
\end{eqnarray}
It follows that the vorticity strength takes the linear form
\begin{equation}
\label{q}
q(x)= -\tilde c_1 - 2 \tilde c_2 x,
\end{equation}
and the constants $c$ and $\delta$ are given by:
\begin{eqnarray}\label{cflat}
c&=& -\tilde c_1^3 + \tilde c_1 \tilde c_3 - \tilde c_2 \tilde c_4,\\ 
\label{delflat}
\delta&=& 3\tilde c_1^2-2\tilde c_3 .
\end{eqnarray}
Finally, the Ricci scalar of the two-dimensional base space is given by
\begin{equation}
\label{Rbaseequil}
\hat R =-2 \left(\tilde c_3 + 6\tilde c_2 x(\tilde c_1 + \tilde c_2 x) \right), 
\end{equation}
and using \eqref{Rsgen} one can easily find the form of the three-dimensional scalar as well. 
Not all the six parameters $\tilde c_i$ correspond to physical quantities: some of them can be just reabsorbed by change of coordinates. In particular, we set here $\tilde c_0 = 0$ by performing the diffeomorphism $t \to t + p\, y$, with constant $p$, which does not change the form of the metric. 

\boldmath
\subsubsection{Non-vanishing $c_4$} \label{nvanc4}
\unboldmath

To analyse the class with $\tilde c_4\neq 0$, we first perform the further diffeomorphism  $x \to x + s$, with constant $s$, which keeps the form of the metric. 
By tuning the value of $s$ we can set $\tilde c_5$ to zero. Therefore, without loss of generality we can choose:
\begin{eqnarray}\label{bgen}
b(x) &=& \tilde c_1 x + \tilde c_2 x^2, \\
\label{Ggen}
G(x) &=& \tilde c_4 x+ \tilde c_3 x^2+\tilde c_2  x^3 \left( 2\tilde c_1+\tilde c_2 x\right).
\end{eqnarray}
We are thus left with four arbitrary geometric parameters. For consistency we can check that $q(x)$, $c$, $\delta$, $R$ and $\hat{R}$ indeed depend only on these four parameters. Moreover, by performing the change of variables
\begin{equation}
\label{c4change}
x\to \frac{x}{\tilde c_4}\,,\quad y\to \frac{y}{\tilde c_4}\,,\quad t\to \frac{t}{\tilde c_4}\,,
\end{equation}
and defining new variables
\begin{equation}
 c_1 = \frac{\tilde c_1}{\tilde c_4}\,,\quad c_2 =\frac{\tilde c_2}{\tilde c_4^2}\,,\quad
 c_3= \frac{\tilde c_3}{\tilde c_4^2}\,,\quad  c_ 4 =  \tilde c_4 
\end{equation}
we can see that $c_4$ is an overall scaling factor of the metric. Indeed, we have
\begin{eqnarray}\label{bgenfinal}
b(x) &=& c_1  x +c_2  x^2, \\
\label{Ggenfinal}
G(x) &=&  x+ c_3  x^2+c_2   x^3 \left( 2c_1+c_2 x\right),
\end{eqnarray}
which depend now on the three dimensionless parameters $c_1$, $c_2$ and $c_3$.
Using the above variables the metric becomes $c_4^2\mathrm{d}s^2$. Since we are dealing with a conformal theory, we can always choose appropriate units to set $c_4$ to any convenient value and deal with dimensionless quantities only. 

\paragraph{Monopoles: homogeneous spaces}\label{TMG}

Consider the vorticity strength \eqref{q}. The simplest example that can be considered is the one of constant $q$, that is when $c_2 = 0$. We call the corresponding geometries \emph{monopolar geometries}, a terminology that we will justify in the following. 
The two-dimensional Ricci scalar \eqref{Rbaseequil} is in this case constant: $\hat{R}=-2c_3$. This means that the parameter $c_3$ labels the curvature signature of the two-dimensional base space and that without loss of generality (by appropriately choosing $c_4$ and then dropping the overall scale factor), we can set
\begin{equation}
c_3=-\nu=0,\pm 1\,.
\end{equation}
Thus, we are left with one continuous parameter, $c_1$,  which we rename as
\begin{equation}
c_1=-2n\,.
\end{equation}
Moreover, the Cotton--York tensor is proportional to
\begin{equation}
c=2n(\nu+4n^2)\,,
\end{equation}
hence the parameter $n$ determines whether the geometry is conformally flat or not. Note that, apart from the trivial case $n=0$, the space is conformally flat also when $\nu=-1$ and $4n^2=1$ -- we will briefly comment on this issue at the end of Sec. \ref{dipax}. 
The functions $b(x)$ and $G(x)$ take now the form
\begin{equation}\label{TMGbst}
b(x) =-2nx\,,\quad G(x)=x(1+\nu x)\,,
\end{equation}
The form of $G(x)$ motivates the parameterisation
\begin{equation}
x = f^2_\nu (\nicefrac{\sigma}{2})\,,\quad\begin{cases}
f_{1}(\sigma)=
\sin \sigma \\ 
f_{0}(\sigma)=\sigma\\
f_{-1}(\sigma)=\sinh \sigma \end{cases}\quad y=2  \varphi\,,\quad\varphi\in[0,2\pi]\,.
\label{TMGmet}
\end{equation}
Then, the geometries 
\eqref{FH-rot-byG} take the form 
\begin{equation}
\label{FH-rot-by1}
\mathrm{d}s^2 = -\left(\mathrm{d}t + 4 n f^2_\nu(\nicefrac{\sigma}{2})\,\mathrm{d}{\bf \varphi}\right)^2+
\mathrm{d}\sigma^2   +f_\nu^2(\sigma) \mathrm{d}{ \varphi}^2\,,
\end{equation}
which is that of fibrations over 
$S^2,\mathbb{R}^2$ and $H_2$ for $\nu = 1, 0, -1$ respectively. The two-dimensional base spaces are  homogeneous with constant curvature having three Killing vectors; the three-dimensional geometry has in total four Killing vectors. 

These geometries appear at the boundary of asymptotically anti-de Sitter Taub--NUT Einstein spaces with $n$ being the bulk nut charge. They were  analysed in detail years ago as families of three-dimensional geometries possessing 4 isometries \cite{rayPRD80, rebPRD83}. As homogeneous space--times, they are of the Bianchi type IX ({ squashed} $S^3$, here as G\"odel space), II (Heisenberg group) and VIII (elliptically {squashed} $\mathrm{AdS}_3$). The second space is also known as Som--Raychaudhuri \cite{SR68}.

We want now to discuss the presence of dangerous tensors. The velocity one-form is: 
\begin{equation}
\mathrm{u} = - \mathrm{d}t - 4 n f^2_\nu (\nicefrac{\sigma}{2}) \mathrm{d}{\varphi},
\end{equation}
while the vorticity has constant strength:
\begin{equation}
q = 2n.
\end{equation}
Furthermore, the geometric data {ensure} the following structure: 
\begin{eqnarray}\label{RgenTMG}
R_{\mu\nu}\,\mathrm{d}x^\mu\mathrm{d}x^\nu =  \left( \nu + 4n^2 \right) \mathrm{u}^2 + \left(\nu + 2 n^2 
\right)\mathrm{d}s^2
\end{eqnarray}
The above condition implies that all hydrodynamic scalars, vectors and tensors that can be constructed from the Riemann tensor, its covariant derivatives and the covariant derivatives of $\mathrm{u}$ are algebraic. More specifically
\begin{itemize}
\item all hydrodynamic scalars are constants,
\item all hydrodynamic vectors are of the form $k u_\mu$ with constant $k$, and
\item all hydrodynamic tensors are of the form $a u_\mu u_\nu + b g_{\mu\nu} $ with constant $a$ and $b$.
\end{itemize}
This means that there exists no traceless transverse tensor that can correct the hydrodynamic energy--momentum tensor in perfect equilibrium. In other words, there exists no dangerous tensor. Thus, in the case of monopolar geometries it is not possible to know the value of any transport coefficient.

This above result is not surprising. Indeed, we called Papapetrou--Randers configurations given by \eqref{TMGmet} and \eqref{FH-rot-by1} of \emph{monopolar} type because the vorticity is constant,\footnote{Note also that $\mathrm{b}$, as given in \eqref{gen-gaug} and \eqref{TMGbst}, has the same form as the gauge potential of a Dirac monopole on $S^2$, $\mathbb{R}^2$ or $H_2$. This magnetic paradigm can be made more precise -- see e.g. \cite{Gibbons}.} as a consequence of the homogeneous nature of these space--times. In such a highly symmetric kinematical configuration, the fluid dynamics cannot be sensitive to any dissipative or non-dissipative coefficient. This  result provides a guide for the subsequent analysis: to have access to the transport coefficients, we must perturb the geometry away from the homogeneous configuration. The above discussion suggests that this perturbation should be organised as a multipolar expansion: the higher the multipole in the geometry, the richer the spectrum of transport coefficients that can contribute, if non-vanishing, to the global equilibrium state, and that we need to set to zero for perfect fluids. 

Finally, we note that the form of the Cotton--York tensor for monopolar geometries is
\begin{equation}\label{CgenTMG}
C_{\mu\nu}
\mathrm{d}x^\mu\mathrm{d}x^\nu = n (\nu + 4 n^2 )(3 \mathrm{u}^2 + \mathrm{d}s^2).
\end{equation}
The above expression can be combined { with} \eqref{RgenTMG}, giving: 
\begin{equation}
\label{TMGeq}
R_{\mu \nu} -\frac{R}{2}g_{\mu \nu}+\lambda g_{\mu \nu}=\frac{1}{\mu}
C_{\mu \nu}.
\end{equation}
The latter shows that monopolar geometries solve the topologically massive gravity equations \cite{cs} for appropriate constant $\lambda$ and $\mu$. This is not surprising, as it is a known fact that, for example, squashed anti-de-Sitter or three-spheres solve topologically massive gravity  equations \cite{Anninos:2008fx}. However, what is worth stressing here is that requiring a generic Papapetrou--Randers background \eqref{Papa}  to solve \eqref{TMGeq} leads necessarily to a monopolar geometry. The argument goes as follows. Using the expression for the Ricci tensor for Papapetrou--Randers geometries \eqref{Rgen}, the left side of \eqref{TMGeq} reads: 

\begin{equation}\label{TMGgen}
\left(\hat R+ \frac{q^2}{2}-6 \lambda\right)\frac{\mathrm{u}^2}{2} +\left(\frac{q^2}{4}+\lambda\right)\left(2\mathrm{u}^2+\mathrm{d}\ell^2\right)
-\mathrm{u}\, \mathrm{d}x^\rho u^\sigma \eta_{\rho\sigma\mu} \nabla^\mu q.
\end{equation}
As the right side of \eqref{TMGeq} is traceless, so should be the left side. This leads to:
\begin{equation}
\label{TMGtr}
\lambda = \frac{\hat{R}}{6} + \frac{q^2}{12}.
\end{equation}
Equations \eqref{TMGeq} can now be analysed using the expression for the Cotton--York tensor \eqref{Cgen} and \eqref{Ggen} together with  \eqref{TMGtr}. The off-diagonal components $\mathrm{u}\, \mathrm{d}x^\rho$ imply that $q$ must be constant. With this at hand, the rest of the equations are automatically satisfied with: 
\begin{equation}\label{TMGmu}
q = \frac{2\mu}{3}.
\end{equation}
In order to provide the general form of a Papapetrou--Randers metric satisfying \eqref{TMGeq}, we can now combine \eqref{TMGtr} with \eqref{TMGmu}. These lead to the conclusion that all solutions are fibrations over a  two-dimensional space with metric $\mathrm{d}\ell^2$ of constant curvature $\hat R= 6 \lambda-\nicefrac{2\mu^2}{9}$. They are thus homogeneous spaces of either positive ($S^2$), null ($\mathbb{R}^2$) or negative curvature ($H_2$). 

The reader might be puzzled by the present connection with  topologically massive gravity. The $2+1$-dimensional geometries analysed here are not supposed to carry any gravity degree of freedom since they are ultimately designed to serve as holographic boundaries. Hence, the emergence of topologically massive gravity  should not be considered as a sign of dynamics, but rather as a constraint for the algebrisation of the operator $\nabla$, which destroys any potential dangerous tensor. Any perfect-Cotton geometry allowing for such tensors, and thereby probing transport coefficients, will necessarily require a deviation from topologically massive gravity.  

More recently, topologically massive gravity has also attracted attention from the holographic perspective \cite{Anninos:2008fx, Anninos:2011vd}. In these works, the homogeneous solutions appear as $2+1$-dimensional \emph{bulk} backgrounds, whereas in the present work (see also \cite{NewPaper}), they will turn out to be naturally leading to \emph{boundary} geometries. Investigating the interplay between these two viewpoints might be of some relevance, beyond our scope though.

\paragraph{Dipolar geometries: axisymmetric spaces}\label{dipax}

When $c_2\neq 0$, the vorticity is not constant and hence the space ceases to be homogenous. If some symmetry remains, this must be in the form of a space-like Killing vector: therefore, these are axisymmetric spaces. We call such geometries \emph{dipolar geometries}, as their axial symmetry connects them with the gauge potential of electric or magnetic dipoles. 

For simplicity, we start considering a \emph{pure dipolar} geometry, namely a nontrivial conformally  flat metric and see how it is parameterised in terms of $c_1, c_2$ and $c_3$. 
We start from $\mathbb{R}\times S^2$ where we set to one the sphere's radius
\begin{equation}
\label{RxS2}
\mathrm{d}s^2=-\mathrm{d}t^2+\mathrm{d}\vartheta^2+\sin^2\vartheta \mathrm{d}\varphi^2.
\end{equation}
We do then a conformal rescaling by a function $\Omega(\vartheta)$, which preserves the axial symmetry around $\varphi$ -- and the conformal flatness of \eqref{RxS2} \emph{i.e.} the vanishing Cotton tensor:
\begin{equation}
\label{Weyl_resc}
\mathrm{d} s'^2=\Omega^{2}(\vartheta)\,  \mathrm{d} s^2\, .
\end{equation}
The vector field $\partial_t$ is no longer of unit norm.
However, if $\Omega(\vartheta)$ simply corresponds to a rotation,\footnote{Actually to a precession, hence we call this the \emph{precession trick}.} then a new unit-norm time-like Killing vector may exist and describe trajectories of a fluid in equilibrium on the background \eqref{Weyl_resc}. We will show that this is possible for 
\begin{equation}
\label{Omega}
\Omega^{2}(\vartheta)=1-a^2\sin^2\vartheta 
\end{equation}
with $a$ being a constant parameter.
Consider for that the change of coordinates $(t, \vartheta,\varphi)\mapsto(t', \vartheta',\varphi')$
defined as:
\begin{equation}
\label{prec_trick}
t=t'\, ,\quad
\Omega^{2}(\vartheta)=\frac{ \Delta_{\vartheta'} }{\Xi}
\, , \quad \varphi=\varphi'+at'
\end{equation}
with 
\begin{equation}
\label{theta-theta'}
\Delta_{\vartheta'} =  1 - a^2 \cos^2\vartheta'\, , \quad 
 \Xi= 1 - a^2\, .
\end{equation}
The metric \eqref{Weyl_resc} reads now:
\begin{equation}
\label{prec_trick2}
\mathrm{d}s'^2=-
      \left[ \mathrm{d}t' - \frac{a}{\Xi} \sin^2\vartheta'  \mathrm{d}\varphi'\right]^2
      + \frac{\mathrm{d}\vartheta'^2}{\Delta_{\vartheta'}}
  +  \frac{\Delta_{\vartheta'}}{\Xi^2} \sin^2\vartheta' \mathrm{d}\varphi'^2\, .
\end{equation}
Clearly the Killing vector 
\begin{equation}
\partial_{t'}=\partial_t+a\partial_{ \varphi}
\end{equation}
is of unit norm, and its vorticity 
\begin{equation}
\omega= \frac{a}{\Xi}\cos\vartheta' \sin \vartheta' \mathrm{d}\vartheta' \wedge \mathrm{d}\varphi'
\end{equation}
has strength 
\begin{equation}
q=-2a\cos\vartheta'\, .
\end{equation}
Any fluid comoving with $\partial_{t'}$ in the background metric \eqref{prec_trick2} undergoes 
a cyclonic rotation on a squashed\footnote{The spatial metric $\mathrm{d}\ell'^2$ in \eqref{prec_trick2}  describes a squashed two-sphere.} $S^2$. As already stressed, this background metric is conformally flat. It is described by a unique parameter $a$ and this is consistent with the analysis of \cite{Grumiller, Guralnik} for conformally flat $2+1$-dimensional geometries.

Finally, by performing the change of coordinates
\begin{equation}
\label{xyphitheta}
x=\frac{\sin^2\nicefrac{\vartheta'}{2}}{1-a^2}
\,,\quad\,y=2\varphi,
\end{equation}
we can bring the metric \eqref{prec_trick2} into the form \eqref{FH-rot-byG} with
\begin{eqnarray}
b(x)&=&2ax\left(1-(1-a^2)x\right)\,, \\ G(x)&=&x-(1-5a^2)x^2-8a^2(1-a^2)x^3+4a^2(1-a^2)^2x^4.
\end{eqnarray}
It is easy then to read off  the parameters
\begin{equation}
\label{c1c2c3}
c_1=2a\,, c_2=-2a(1-a^2)\,,c_3=5a^2-1.
\end{equation}

We now move to the generalisation of the above to non-conformally flat geometries with $x$-dependent vorticity. These are the \emph{dipolar-monopolar} metrics. In those cases the precession trick mentioned previously does not suffice and one needs to perform the appropriate parameterisations of the $c_i$'s.
Nevertheless, our previous explicit examples serve both as a guiding rule as well as a test for our results. We will present them and spare the reader from the non-illuminating technicalities. By appropriately choosing $c_4$ and dropping the overall scale factor, we can parameterise $c_1$, $c_2$ and $c_3$ by the charge $n$ and the angular momentum $a$, without loss of generality. These parameterisations will depend on the topology captured in  $\nu$. As already quoted, from the holographic analysis presented in Sec. \ref{hol}, it will become clear that $n$ is the bulk nut charge.

\paragraph*{Spherical ($\nu=1$)}    The relation between $a$ and $n$ and the three geometric parameters is given by:
\begin{eqnarray}\label{cdipsp}
c_1 &=& 2  (a-n), \nonumber\\
c_2 &=& 2 a  (-1 + a^2  - 4 a n  ), \nonumber\\
c_3 &=&  -1 + 5 a^2  - 12 an.
\end{eqnarray}
We also perform the following coordinate transformations:
\begin{eqnarray}\label{transfsp}
x &=& \kappa \sin^2 \nicefrac{\vartheta}{2}, \nonumber\\
y &=& \lambda \varphi ,
\end{eqnarray}
with
\begin{equation}\label{klsp}
\kappa = \frac{1}{1 + a(4 n-a)}, \quad \lambda = \frac{2}{ \kappa\, \Xi} \quad \text{and}\quad \Xi =1 -  a^2.
\end{equation}
The two-dimensional base space in the metric \eqref{FH-rot-byG} takes then the form: 
\begin{equation}
\label{dlSKTN}
\mathrm{d}\ell^2=
\frac{\mathrm{d}\vartheta^2}{\Delta_\vartheta}+\frac{\sin^2\vartheta\Delta_\vartheta}{\Xi^2}\mathrm{d}\varphi^2
\end{equation}
with
\begin{equation}
\label{dataKTN}
\Delta_\vartheta=1+a\cos\vartheta(4n-a\cos\vartheta).
\end{equation}
The coordinates range as $\vartheta\in [0,\pi]$ and $\varphi\in [0,2\pi]$. The full $2+1$-dimensional metric is of the Papapetrou--Randers form: $\mathrm{d}s^2 = -\mathrm{u}^2 + \mathrm{d}\ell^2$. The velocity field takes the form 
\begin{equation}
\label{bSKTN}
\mathrm{u}=-\mathrm{d}t+  b (\vartheta)\mathrm{d}\varphi\,,\quad b(\vartheta) =  \frac{2 (a - 2 n + a \cos\vartheta)}{\Xi}\sin^2\nicefrac{\vartheta}{2}.\
\end{equation}
The scalar vorticity strength is given by
\begin{equation}
\label{qSKTN}
q=2\left(n-a\cos \vartheta\right),
\end{equation}
while the constant $c$ appearing in the Cotton--York tensor is
\begin{equation}
 c  = 2  n (1 - a^2  + 4  n^2).
\end{equation} 
The base space \eqref{dlSKTN} is a squashed $S^2$. The vorticity \eqref{qSKTN} has two terms: the constant monopole and the dipole. It is maximal on the northern ($\vartheta = 0$) and southern ($\vartheta=\pi$) poles and is vanishing on the equator ($\vartheta=\nicefrac{\pi}{2}$). Note also that in the limit {$c_2=0$} we {recover} the homogeneous metric case for $\nu = 1$.

\paragraph*{Flat ($\nu=0$)}         
The new parameters $a$ and $n$ are now defined as follows:
\begin{eqnarray}\label{cdipfl}
c_1 &=& 2  (a-n), \nonumber\\
c_2 &=& 2 a^2  (a -4n), \nonumber\\
c_3 &=& a (5 a - 12 n).
\end{eqnarray}
Let us now do the following coordinate transformations:
\begin{eqnarray}\label{transfpl}
x &=& \kappa(\nicefrac{\sigma}{2})^2, \nonumber\\
y &=& \lambda \varphi ,
\end{eqnarray}
with
\begin{equation}\label{klpl}
\kappa = {1}, \quad \lambda = 2.
\end{equation}
With these transformations the two-dimensional base space in the metric \eqref{FH-rot-byG} takes the form of squashed $\mathbb{R}^2$: 
\begin{equation}
\label{dlFKTN}
\mathrm{d}\ell^2=
\frac{\mathrm{d}\sigma^2}{\Delta_\sigma}+\sigma^2\Delta_\sigma\mathrm{d}\varphi^2
\end{equation}
with
\begin{equation}
\label{dataFKTN}
\Delta_\sigma={\frac{1}{16}}(2 + a^2  \sigma^2) (8 - 24 a n   \sigma^2 + 
    a^4  \sigma^4 - 8 a^3 n   \sigma^4 + 
    2 a^2  \sigma^2 (3 + 8  n^2 \sigma^2)).
\end{equation}
The coordinates range as $\sigma \in \mathbb{R}_+$ and $\varphi\in [0,2\pi]$. The full $2+1$-dimensional metric is $\mathrm{d}s^2 = -\mathrm{u}^2 + \mathrm{d}\ell^2$, where the velocity field takes the form 
\begin{equation}
\label{bFKTN}
\mathrm{u}=-\mathrm{d}t+  b (\sigma)\mathrm{d}\varphi\,,\quad b (\sigma)=\frac{\sigma^2}{4} \left(4 (a - n) + a^2  (a - 4 n) \sigma^2\right).
\end{equation}
The scalar vorticity is then given by
\begin{equation}
\label{qFKTN}
q=(n-a)\left(2 + a^2  \sigma^2\right), 
\end{equation}
while the constant $c$ appearing in Cotton--York tensor is:
\begin{equation}
 c  = 2  n (-a^2 + 4 n^2).
\end{equation} 

\boldmath
\paragraph*{Hyperbolic case ($\nu=-1$)}           
\unboldmath
This case is very similar to the spherical one, with trigonometric functions traded for hyperbolic ones. We define $a$ and $n$ using:
\begin{eqnarray}\label{cdiphy}
c_1 &=& 2  (a-n), \nonumber\\
c_2 &=& 2 a  (1 + a^2  - 4 a n  ), \nonumber\\
c_3 &=&  1 + 5 a^2  - 12 an.
\end{eqnarray}
The appropriate coordinate transformations are:
\begin{eqnarray}\label{transfhy}
x &=& \kappa \sinh^2 \nicefrac{\sigma}{2}, \nonumber\\
y &=& \lambda \varphi ,
\end{eqnarray}
with
\begin{equation}\label{klhy}
\kappa = \frac{1}{1 - a(4n-a) }, \quad \lambda = \frac{2}{ \kappa Z} \quad \text{and}\quad Z = 1 +  a^2.
\end{equation}
With these transformations the two-dimensional base space in the metric \eqref{FH-rot-byG} takes the form of squashed $H_2$: 
\begin{equation}
\label{dlHKTN}
\mathrm{d}\ell^2=
\frac{\mathrm{d}\sigma^2}{\Delta_\sigma}+\frac{\sinh^2\sigma\Delta_\sigma}{\Xi^2}\mathrm{d}\varphi^2
\end{equation}
with
\begin{equation}
\label{dataHKTN}
\Delta_\sigma=1-a\cosh\sigma(4n-a\cosh\sigma).
\end{equation}
The coordinates range as $\sigma\in \mathbb{R}_+$ and $\varphi\in [0,2\pi]$. In the full $2+1$-dimensional metric $\mathrm{d}s^2 = -\mathrm{u}^2 + \mathrm{d}\ell^2$, the velocity field takes the form 
\begin{equation}
\label{bHKTN}
\mathrm{u}=-\mathrm{d}t+  b (\sigma)\mathrm{d}\varphi\,,\quad b(\sigma) = \frac{2 (a - 2 n + a \cosh\sigma)}{Z}\sinh^2\nicefrac{\sigma}{2}.
\end{equation}
The scalar vorticity is
\begin{equation}
\label{qHKTN}
q=2\left(n-a\cosh \sigma\right),
\end{equation}
while the constant $c$ appearing in the Cotton--York tensor is
\begin{equation}
 c  = 2  n (-1 - a^2  + 4  n^2).
\end{equation}

\paragraph*{Uniform parameterisation} It is possible to use a uniform notation to include the three different cases:
\begin{eqnarray}\label{cdipgen}
c_1 &=& 2  (a-n), \nonumber\\
c_2 &=& 2 a  (-\nu + a^2  - 4 a n  ), \nonumber\\
c_3 &=&  -\nu + 5 a^2  - 12 an.
\end{eqnarray}
The general coordinate transformations are:
\begin{eqnarray}\label{transfgen}
x &=& \kappa f^2_\nu (\nicefrac{\theta}{2}), \nonumber\\
y &=& \lambda \varphi ,
\end{eqnarray}
with $f_\nu$ as in \eqref{TMGmet}, and
\begin{equation}\label{klgen}
\kappa = \frac{1}{1 + \nu a (4n-a)}, \quad \lambda = \frac{2}{ \kappa Z_\nu} \quad \text{and}\quad Z_\nu = 1 - \nu  a^2.
\end{equation}
The constant $c$ appearing in Cotton--York tensor takes the form:
\begin{equation}
\label{cgen}
 c  = 2  n (\nu - a^2  + 4  n^2).
\end{equation} 

Before moving to the general case $c_4 = 0$, a comment is in order here. One observes from 
\eqref{cgen} that the Cotton tensor of the monopole--dipole $2+1$ geometries may vanish in two distinct instances. The first is when the charge $n$ itself vanishes, which corresponds to the absence of the monopolar component. The second occurs when
\begin{equation}
\label{cvan}
\nu - a^2  + 4  n^2=0.
\end{equation} 
For vanishing $a$, only the case $\nu=-1$ is relevant:\footnote{For $\nu=0$, we recover again $n=0$ and the $2+1$ geometry is $\mathbb{R}^{2+1}$, whereas $\nu=1$ requires $n=\pm\nicefrac{i}{2}$, which produces a signature flip to $(-++)\to (+++)$, with geometry $S^3$.}
$n=\pm\nicefrac{1}{2}$ and geometry $\text{AdS}_3$. For non-vanishing $a$, we obtain a conformally flat, non-homogeneous deformation of the $n$-squashed\footnote{Again for $\nu=1$, there is a signature flip, unless $a^2>1$. From the bulk perspective, where $a$ is the rigid angular velocity (see Sec. \ref{hol}), this corresponds to an ultra-spinning black hole and is an unstable situation \cite{Hawking:1998kw}.}  $S^3, \mathbb{R}^{2+1}$ or $\text{AdS}_3$.

\boldmath
\subsubsection{Vanishing $c_4$} \label{dipvanc4}
\unboldmath

When the parameter $\tilde c_4\equiv c_4$ is vanishing, it is not possible to perform the change of variables \eqref{c4change} and thus we have a different class of metrics. We are left with the parameters $\tilde c_1 \equiv c_1$, $\tilde c_2 \equiv c_2$, $\tilde c_3 \equiv c_3$ and $\tilde c_5 \equiv c_5$. We decide not to set to zero the latter in order to avoid a possible metric singularity (see \eqref{G}) when $c_2 = c_3 = 0$. The boundary metric is in this case given by
\begin{equation}
\label{c4van}
\begin{split}
b(x) &= c_1 x + c_2 x^2, \\
G(x) &=  c_5 + c_3 x^2 + c_2 x^3 (2 c_1 + c_2 x),
\end{split}
\end{equation}
with
\begin{equation}
c=c_1\left(c_3-c_1^2\right).
\end{equation}
For the flat horizon case $c_3 = 0$, this class of metrics appears 
as boundary of Einstein solutions studied in \cite{Ortin}. When $c_2 = 0$ we have a homogeneous geometry and what we concluded on transport coefficients for the case before is still valid: it is not possible to constrain any of them holographically, because the corresponding tensors vanish kinematically. 

As in the previous situation, the boundary geometries at hand can be conformally flat. This occurs either when $c_1$ vanishes, or when 
\begin{equation}\label{cvanc4van}
c_3 = c_1^2.
\end{equation}

\section{The bulk duals of perfect equilibrium}\label{hol}

\subsection{Generic bulk reconstruction}
When the boundary geometry is of the perfect-Cotton type and the boundary stress tensor is that of a fluid in perfect equilibrium, the bulk solution can be exactly determined. This is highly non-trivial because it generally involves an infinite resummation \emph{i.e.} starting from the boundary data and working our way to the bulk. 

The apparent resummability of the boundary data discussed above into exact bulk geometries is remarkable, but not too surprising. An early simple example was given in \cite{SS} where it was shown that setting the boundary energy--momentum tensor to zero and starting with a conformally flat boundary metric, one can find the (conformally flat) bulk solution resuming the Fefferman--Graham series --  in that case the resummation involved just a few terms. 

The next non-trivial example was presented in \cite{Mansi:2008bs} (see also 
\cite{deHaro:2008gp}). There, it was shown that in Euclidean signature, imposing the condition 
\begin{equation}
C_{\mu\nu} =\pm 8\pi G_{\mathrm{N}}T_{\mu\nu}
\end{equation}
is exactly equivalent to the (anti)-self duality of the bulk Weyl tensor,\footnote{More generally, the boundary Cotton tensor is an asymptotic component of the bulk Weyl tensor (see e.g. Eq. (2.8) of \cite{Mansi:2008br}). However, a non-vanishing Weyl does not necessarily imply a non-vanishing Cotton, as for example in the Kerr--AdS$_4$ case.  A non-vanishing Cotton, on the other hand, requires the Weyl be non-zero. Non-cyclonic vorticity on the boundary requires precisely non-zero Cotton, as we discuss in Sec. \ref{pCg}. 
This was unambiguously stated in \cite{Leigh:2011au, LPP2, NewPaper} in relation with the nut charge, the latter being encapsulated in the Weyl component  $\Psi_2$ (see Griffiths and Podolsk\'y  p. 215 \cite{GP}, and for applications in fluid/gravity correspondence \cite{Eling:2013sna}). Clearly, the structure of the perfect Cotton puts constraints on the bulk Weyl tensor.
} 
hence it leads to conformal (anti)-self dual solutions. This property was also discussed in \cite{FG}. 
In fact, prior to the advent of holography, the problem of \emph{filling-in} Berger three-spheres with (potentially conformal self-dual) Einstein metrics was addressing the same issues, in different terms and in a Euclidean framework \cite{LeBrun82, Pedepoon90, Tod90, Tod94, Hitchin95}.\footnote{Euclidean four-dimensional conformal (anti)-self dual Einstein manifolds are known as \emph{quaternionic}  and include spaces such as Fubini--Study or {Calderbank}--Pedersen.} However, in these cases it is not clear whether the boundary theory describes a hydrodynamical system. 
Here we study a particular extension of the (anti)-self dual boundary condition of \cite{Mansi:2008bs}, which is Lorentzian and reads:
\begin{equation}
\label{TtC}
C_{\mu\nu} = \chi T_{\mu\nu}\,,\quad\,\chi=\frac{c}{\varepsilon}\, ,
\end{equation}
with both $T_{\mu\nu}$ and $C_{\mu\nu}$ having the \emph{perfect-fluid} form and $\chi\neq 8\pi G_{\mathrm{N}}$ generically.

Our main observation is that to the choice (\ref{TtC}) for the boundary data corresponds the following \emph{exact} bulk Einstein metric in Eddington--Finkelstein coordinates (where $g_{rr}=0$ and $g_{r\mu} = - u_\mu$):
\begin{eqnarray}\label{4d.CpropT}
\mathrm{d}s^2 &=& - 2 \mathrm{u} \left( \mathrm{d}r -\frac{1}{2}\mathrm{d}x^\rho u^\sigma \eta_{\rho\sigma\mu}\nabla^\mu q\right)+ \rho^2 \mathrm{d}\ell^2 \nonumber \\
 &&- 
\left(r^2+\frac{\delta}{2}-\frac{q^2}{4} 
-\frac{1}{\rho^2}\left(2Mr+ \frac{ q c}{2}\right) \right)\mathrm{u}^2 ,
\end{eqnarray}
with
\begin{equation}
\label{rho}
\rho^2 = r^2 + \frac{q^2}{4}.
\end{equation}
The metric above is manifestly covariant with respect to the boundary metric. Taking the limit $r\rightarrow \infty$ it is easy to see that the boundary geometry is indeed the general stationary Papapetrou--Randers metric in \eqref{Papa} with
\begin{equation}
\mathrm{u} = -\mathrm{d}t + b\, \mathrm{d}y.
\end{equation}

The various quantities appearing in \eqref{4d.CpropT} (like $\delta, q, c$) satisfy Eqs. \eqref{cotequileq1}, \eqref{cotequileq2} and \eqref{Requil}, and this guarantees that Einstein's equations are satisfied. Performing the Fefferman--Graham expansion of \eqref{4d.CpropT}  we indeed recover  the perfect form  of the boundary energy--momentum tensor with
\begin{equation}
\label{eden}
\varepsilon = \frac{M}{4\pi G_{\mathrm{N}}},
\end{equation}
where $G_{\mathrm{N}}$ is the four-dimensional Newton's constant.
The corresponding holographic  fluid has velocity field $\mathrm{u}$, vorticity strength $q$ and behaves like a perfect fluid.

In the choice of gauge given by \eqref{gen-2D} and \eqref{gen-gaug}, the bulk metric \eqref{4d.CpropT} takes the form:
\begin{eqnarray}\label{bulkmetric1}
\mathrm{d}s^2 &=& - 2 \mathrm{u} \left( \mathrm{d}r- \frac{1}{2} \left( \mathrm{d}y \frac{B}{A} \partial_x q - \mathrm{d}x \frac{A}{B} \partial_y q \right)\right) + \rho^2 \mathrm{d}\ell^2 \nonumber \\
 &&- 
\left(r^2+\frac{\delta}{2}-\frac{q^2}{4}
-\frac{1}{\rho^2}\left(2Mr+ \frac{ q c}{2 }\right) \right)\mathrm{u}^2 ,
\end{eqnarray}
where $q$ is as in \eqref{gen-q}. Note $\delta$ and $c$ can be readily obtained from $q$, $A$ and $B$ using \eqref{Requil} and \eqref{cotequileq1} respectively.

It is clear from the explicit form of the bulk spacetime metric \eqref{4d.CpropT} that the metric has a curvature singularity when $\rho^2 = 0$. The locus of this singularity is at :
\begin{equation}\label{sing}
r = 0, \quad q(x,y) = 0.
\end{equation}
However, we will find cases where $\rho^2$ never vanishes because $q^2$ never becomes zero. In such cases, the bulk geometries have no curvature singularities, but they might have regions with closed time-like curves.
  
{Although the  Killing vector $\partial_t$}
is of unit norm at the boundary coinciding with the velocity vector of the boundary fluid, it's norm is not any more unity in the interior. {In particular,} the Killing vector becomes null at the ergosphere $r= R(x)$ where:
\begin{equation}
r^2+\frac{\delta}{2}-\frac{q^2}{4}
-\frac{1}{\rho^2}\left(2Mr+ \frac{ q c}{2 }\right) = 0.
\end{equation}
Beyond the ergosphere no observer can remain stationary, and hence he experiences frame dragging,  as $\partial_t$ becomes space-like. 

Before closing this section, a last comment is in order, regarding the exactness of the bulk solution  
\eqref{4d.CpropT}--\eqref{rho}, obtained by uplifting $2+1$-dimensional perfect boundary data \emph{i.e.} perfect energy--momentum tensor \eqref{T0} and perfect-Cotton boundary geometry \eqref{perfCott}.

The Fefferman--Graham expansion, quoted previously as a way to organise the boundary (holographic) data, is controlled  by the inverse of the radial coordinate $\nicefrac{1}{r}$.   An alternative expansion has been proposed in  \cite{Hubeny, Rangamani:2009xk}. This is a derivative expansion (long wavelength approximation) that modifies order by order the bulk geometry, all the way from the horizon to the asymptotic region.  It has been investigated from various perspectives  in bulk dimension greater than 4. In the course of this investigation, it was observed \cite{Bhattacharyya:20082, Bhattacharyya:2008ji}  that for AdS--Kerr geometries, at least in 4 and 5 dimensions,  the derivative expansion obtained with a perfect energy--momentum tensor and the Kerr boundary geometry,  turns out to reproduce exactly the bulk geometry, already at first order, modulo an appropriate resummation that amounts to redefining the radial coordinate. 

Lately, it has been shown \cite{NewPaper} that the above observation holds for the Taub--NUT geometry in 4 dimensions provided the quoted derivative expansion includes a higher-order term
involving the Cotton--York tensor of the boundary geometry. The derivative expansion up to that order reads: 
\begin{equation}
\label{papaef}
\mathrm{d}s^2 =
-2\mathrm{u}\mathrm{d}r+r^2\mathrm{d}s^2_{\mathrm{bry.}}+\Sigma_{\mu\nu} 
\mathrm{d}x^\mu\mathrm{d}x^\nu
+ \frac{\mathrm{u}^2}{\rho^2}  \left(2Mr+\frac{1}{2}u^\lambda C_{\lambda \mu}\eta^{\mu\nu\sigma}\omega_{\nu\sigma}\right),
\end{equation}
where all the quantities refer to the boundary metric $\mathrm{d}s^2_{\mathrm{bry.}}$ of the Papapetrou--Randers type \eqref{Papa}, and $\mathrm{u}$ is the velocity field of the fluid that enters the perfect energy--momentum tensor \eqref{T00}, whose energy density is related to $M$ according to \eqref{eden}. Furthermore, 
\begin{eqnarray}
\Sigma_{\mu\nu} 
\mathrm{d}x^\mu\mathrm{d}x^\nu&=&-2\mathrm{u}\nabla_\nu \omega^\nu_{\hphantom{\nu}\mu}\mathrm{d}x^\mu- \omega_\mu^{\hphantom{\mu}\lambda} \omega^{\vphantom{\lambda}}_{\lambda\nu}\mathrm{d}x^\mu\mathrm{d}x^\nu
-\mathrm{u}^2\frac{R}{2},\\ \label{rho2}
\rho^2&=& r^2 +\frac{1}{2} \omega_{\mu\nu} \omega^{\mu\nu},
\end{eqnarray}
where, as usual $\omega_{\mu\nu}$ are the components of the vorticity and $R$ the curvature 
of the boundary geometry.  Metric \eqref{papaef} is the expansion stopped at the fourth derivative of the velocity field (the Cotton--York counts for three derivatives).\footnote{Strictly speaking, the redefinition $\rho(r)$ \eqref{rho2} accounts for a full series with respect to the vorticity, \emph{i.e.} contains terms up to infinite velocity derivatives.} It was shown to be exact for the Taub--NUT boundary in \cite{NewPaper} -- as well as for Kerr whose boundary has vanishing Cotton. 

Metric \eqref{papaef} coincides precisely with \eqref{4d.CpropT} for perfect-Cotton boundary geometries. This identification explains a posteriori the observation of \cite{Bhattacharyya:20082, Bhattacharyya:2008ji} about the exactness of the limited derivative expansion (up to the redefinition $\rho(r)$), and generalises it to all perfect-Cotton geometries with perfect-fluid energy--momentum tensor.  It raises also the question whether similar properties hold in higher dimensions, following the already observed exactness of the lowest term for Kerr. In particular one may wonder what replaces the perfect-Cotton geometry in higher dimensions, where there is no Cotton--York tensor. As we stressed, the bulk gravitational duality is a guiding principle that translates precisely to the boundary Cotton/energy--momentum relationship used in this paper. A similar principle is not available in every dimension and we expect only a limited number of cases where the observation made in \cite{Bhattacharyya:20082, Bhattacharyya:2008ji} about Kerr could be generalised to more general Einstein spaces.

\subsection{Absence of naked singularities}

{We will now} show explicitly that for all perfect-Cotton geometries in this class, the bulk geometries have no naked singularities for appropriate range of values of the black hole mass. Our general solutions will be labeled by three parameters - namely the angular momentum $a$, the nut charge $n$ and the black hole mass $M$. This will cover all known solutions and also give us some new ones, as will be shown explicitly later in Appendix \ref{solutions}.

In order to analyse the bulk geometry we need to know the boundary geometry explicitly. In the previous section, we have been able to find all the perfect-Cotton geometries. These geometries,  {which systematically possess an extra spatial Killing vector,}  are given by \eqref{FH-rot-byG}, \eqref{bgenfinal} and \eqref{Ggenfinal}, and are labelled by three continuously variable parameters $c_1$, $c_2$ and $c_3$. We have shown that without loss of generality, we can rewrite these parameters in terms of the angular momentum $a$, the nut charge $n$ and a discrete variable $\nu$ as in Eq. (\ref{cdipgen}).

The holographic bulk dual \eqref{4d.CpropT} for perfect equilibrium in these general boundary geometries then reads:
\begin{eqnarray}
\mathrm{d}s^2 &=& - 2 \mathrm{u} \left( \mathrm{d}r- \frac{G}{2} \partial_x q  \, \mathrm{d}y  \right) + \rho^2 \Big(\frac{\mathrm{d}x^2}{G}+ G\mathrm{d}y^2 \Big)\nonumber \\
 &&- 
\left(r^2+\frac{\delta}{2}-\frac{q^2}{4}
-\frac{1}{\rho^2}\left(2Mr+ \frac{ q c}{2 }\right) \right)\mathrm{u}^2 ,
\label{4d.CpropT-iso}
\end{eqnarray}
where $\mathrm{u} =-\mathrm{d}t+b \mathrm{d}y$, and $b$ and $G$ are determined by three geometric $c_1$, $c_2$ and $c_3$ as in \eqref{bgenfinal} and \eqref{Ggenfinal}. Therefore $q$, $c$ and $\delta $ are as in \eqref{q}, \eqref{cflat} and \eqref{delflat} respectively. 

It is convenient for the subsequent analysis to move from Eddington--Finkelstein to Boyer--{Lindqvist} coordinates. These Boyer--{Lindqvist} coordinates make the location of the horizon manifest. These are the analogue of Schwarzschild coordinates in presence of an axial symmetry. In the case when the geometric parameter $c_4$ is non-vanishing, the transition to Boyer--{Lindqvist} coordinates can be achieved via the following coordinate transformations:
\begin{eqnarray}
\mathrm{d}\tilde{t} &=& \mathrm{d}t -\frac{4(c_1^2 + 4 r^2) }{3c_1^4 + 8 c_1 c_2  - 4c_1^2 (c_3 + 6r^2)+ 16 r(2M + c_3 r-r^3)}\mathrm{d}r, \\
\mathrm{d}\tilde{y} &=& \mathrm{d}y + \frac{16 c_2}{3c_1^4 + 8 c_1 c_2  - 4c_1^2 (c_3 + 6r^2)+ 16 r(2M + c_3 r-r^3)}\mathrm{d}r.
\end{eqnarray}
Note {that} even after changing $t,y$ to $\tilde{t}, \tilde{y}$, the boundary metric still remains the same - the difference between the old and new coordinates die off asymptotically.

After these transformations the bulk metric takes the form (we replace $\tilde{r}$ and $\tilde{y}$ with $r$ and $y$):
\begin{equation}\label{bulkmetricBG}
\mathrm d s^2 = \frac{\rho^2}{\Delta_r}\mathrm{d}r^2 
- \frac{\Delta_r}{\rho^2}\left(\mathrm{d}t + \beta\mathrm{d}y \right)^2
+\frac{\rho^2}{\Delta_x}\mathrm{d}x^2 +  \frac{\Delta_x}{\rho^2}\left(c_2 \mathrm{d}t - \alpha \mathrm{d}y \right)^2,
\end{equation}
where
\begin{eqnarray}
\rho^2 &=& r^2 + \frac{q^2}{4} \, =\, r^2 + \frac{(c_1 + 2 c_2x)^2}{4}, \\
\Delta_r &=& - \frac{1}{16}\left( 3 c_1^4 + 8 c_1 c_2  - 4 c_1^2(c_3 + 6 r^2)+ 16 r (2M + c_3 r - r^3)\right), \\
\Delta_x &=& G \,\, = \, \, x + c_3 x^2 + 2 c_1 c_2 x^3 + c_2^2 x^4, \\
\alpha &=& -\frac{1}{4} \left(c_1^2 + 4 r^2\right), \\
\beta &=& -b \,\, = \, \, - c_1x - c_2 x^2.
\end{eqnarray}
{The} coordinates $r$ and $x$ do not change as we transform from Eddington--Finkelstein to Boyer--{Lindqvist} coordinates. Therefore $\rho^2$ is exactly the same as before. Also note that $\Delta_r$ and $\alpha$ are functions of $r$ only, while $\Delta_x$ and $\beta$ are functions of $x$ only.

It is easy to see that the horizons are {located} at $r = r_*$ where:
\begin{equation}\label{horizon}
\Delta_r(r = r_*) = 0, \quad \text{with} \quad r_* > 0.
\end{equation}
At most we can have four horizons. These horizon(s) should clothe the curvature singularity located at $\rho^2 = 0$ or equivalently at:
\begin{equation}
r = 0, \quad x = - \frac{c_1}{2c_2}.  
\end{equation}
It is not hard to see that for fixed values of the geometric parameters $c_1$, $c_2$ and $c_3$, there exists a positive definite solution to Eq. (\ref{horizon}) for an appropriate range of the black hole mass $M$. Hence the  curvature singularity is not naked.

Clearly we have only two Killing vectors generically - namely $\partial_t$ and $\partial_y$. Each horizon $r = r_*$ is generated by the Killing vector:
\begin{equation}
\partial_t + \Omega_{\rm{H}} (r_*) \partial_y.
\end{equation}
which is an appropriate linear combination of the two Killing vectors. $\Omega_{\rm{H}}(r_*)$ is a constant given by:
\begin{equation}
\Omega_{\rm{H}}(r_*) = \frac{c_2}{\alpha(r_*)}
\end{equation}
and is the rigid velocity of the corresponding horizon.

The bulk geometry can have at most four ergospheres where the Killing vector $\partial_t$ becomes null. These are given by $r = R(x)$ where $R(x)$ is a solution of:
\begin{equation}
g_{tt} = 0, \quad \text{i.e.} \quad
\Delta_r = c_2^2 G.
\end{equation}

We have seen in Section \ref{dipax} that the geometric structure of the boundary geometries is better revealed as fibrations over squashed $S^2$, $\mathbb{R}^2$ or $H_2$ if we do a further coordinate transformation in $x$ and $y$. We will do the same coordinate transformations given by \eqref{transfgen} in the bulk metric separately for $\nu= 1, 0, -1$. We will also need to exchange parameters $c_1$, $c_2$ and $c_3$ with $a$, $n$ and $\nu$ using \eqref{cdipgen}. Note in these coordinate transformations the radial coordinate $r$ and the time coordinate $t$ do not change, while the spatial coordinates $x$ and $y$ transform only as functions of themselves. This preserves the Boyer--Lindqvist form of the metric \eqref{bulkmetricBG}. We can apply the same strategy to locate the horizon(s) and the ergosphere(s). 

The advantage of doing these coordinate transformations is that for $\nu = 1, 0, -1$ we will see that the horizon will be a squashed $S^2$, $\mathbb{R}^2$ and $H_2$ respectively. The metrics are given explicitly in Appendix \ref{solutions}, where we will also show that we recover all known rotating black hole solutions for which the horizons will be squashed $S^2$ or $H_2$. As far as we are aware of the literature, the case of squashed $\mathbb{R}^2$ horizon \eqref{fh-KNT} is novel.

For the case of vanishing $c_4$, we can similarly proceed to change coordinates {and} bring the bulk metric to Boyer--{Lindqvist} form. The details are presented in Appendix \ref{solutions}, Eq. \eqref{bulkmetricBGa}. Except for the special case \eqref{ortin}, all such solutions in this class will be novel as far as we are aware of the literature.

Interestingly when $c_2=0$, $\rho^2> c_1^2/4$, hence it never vanishes. Therefore the bulk geometry has no curvature singularity. In terms of $a$, $n$ and $\nu$, this happens when
\begin{itemize}
\item for $\nu = 1$: $n>a$;
\item for $\nu = 0$: $n>a$ or $n<\nicefrac{a}{4}$;
\item for $\nu = -1$: $n<a$ or $\vert n\vert \leq \nicefrac{1}{2}$.
\end{itemize}
In such cases horizon(s) may exist, but in absence of a curvature singularity, it is not necessary for the horizon to exist in order that the solution is a good solution.

\subsection{Comment on the rigidity theorem}

As we have shown in previous sections, the perfect-Cotton condition forces the geometry to have at least an additional spatial isometry. This is consistent with the rigidity theorem in $3+1$-dimensions which requires all stationary black hole solutions in flat spacetime to have an axial symmetry. However, as far as we are aware, it is not known if this theorem is valid for $3+1$-dimensional asymptotically AdS stationary black holes. Our results appear thus as an indirect and somehow unexpected hint in favor of the rigidity theorem beyond asymptotically flat spacetimes.

\subsection{Black hole uniqueness from perfect fluidity}

In the generic boundary geometries discussed here, there is a unique time-like Killing vector of unit norm. Physically this corresponds to the fluid velocity field of the perfect-equilibrium state at the boundary.  

The basic observation is that if all stationary black holes in anti-de Sitter space  are dual to perfect-equilibrium states in the CFT, then they are generically unique and are labeled by the mass $M$ for a fixed boundary geometry. The uniqueness is simply a consequence of the fact that there is a unique solution of fluid mechanics, which is in perfect equilibrium in the boundary geometry, as given by Eq. (\ref{Euler0K}).

One may of course wonder why and how geometric parameters of the black hole are related to global thermodynamic parameters describing a perfect-equilibrium state. This question is relevant because the local equation of state is independent of the geometry and is an intrinsic property of the microscopic theory. In fact in a CFT it is simply $\varepsilon = 2p$ (which is also imposed as a constraint of Einstein's equations in the bulk). Nevertheless, global thermodynamics describing the black-hole geometry will depend on the choice of boundary geometry. The thermodynamic charges can be constructed by suitably integrating $T^{\mu\nu}$ over the boundary manifold \cite{Clarkson:2002uj}. In fact some of the geometric parameters will be related to conserved charges -- like $a$ will be related to the angular momentum. The intrinsic variables -- namely the temperature $T$ and the angular velocity $\Omega$ -- can be determined either by using thermodynamic identities or by using the properties of the outermost horizon.

For certain values of parameters we will get instances where there will be extra isometries (like boosts in flat space) which are broken by the perfect-equilibrium fluid configuration. In that case we can generate new solutions by applying these isometries on the fluid configuration (like boosting $\mathrm{u}$). For each such isometry, we will have an additional parameter labeling these black hole solutions (as in the case of boosted black branes).

In the case of space--times with an additional spatial isometry  { as dictated by the boundary perfect-Cotton condition}, the black hole solutions are uniquely described by four parameters, namely $M$ and the three geometric parameters $a$, $n$ and $\nu$ for generic values. Since the perfect-equilibrium solution preserves the additional spatial isometries, the latter cannot be used to generate any new solution.

\section{Constraints on transport coefficients}\label{trco}

In the previous section, we have shown that we can find exact black-hole solutions corresponding to perfect equilibrium of the dual field theory in perfect-Cotton boundary geometries.  From the perspective of the boundary fluid dynamics, by construction, the energy--momentum tensor is \emph{exactly} of the perfect type. Thus any dangerous tensor that this deformed boundary may have, will necessarily couple to vanishing transport coefficients. This gives non-trivial information about strongly coupled holographic conformal fluids in the classical gravity approximation.

We will explicitly show here that exact black-hole solutions indeed imply holographic fluids at strong coupling and in the classical gravity approximation have infinitely many vanishing transport coefficients. On a cautionary note, using perfect-Cotton geometries at the boundary, we will not be able to constrain all transport coefficients. This is because many Weyl-covariant, traceless and transverse tensors will vanish kinematically. We will need to know all possible holographic perfect geometries, or equivalently all exact black-hole solutions with regular horizons, in order to know which transport coefficients vanish in three-dimensional conformal holographic fluids at strong coupling and in the classical gravity approximation. This is possibly not true and it's investigation is also beyond the scope of the present work.

We should also keep in mind that small anti-de Sitter black holes can develop instabilities. Our subsequent conclusions on transport coefficients, hold provided we are in the correct range of parameters. Being concrete on this issue requires to handle the thermodynamic properties and the phase diagram of the black holes at hand, which is a  difficult task in the presence of nut charges. We will leave this analysis for the future and assume for the present being in the appropriate regime for our results to be valid.

We have seen in Sec. \ref{TMG} that a class of perfect-Cotton geometries corresponding to homogeneous backgrounds have no dangerous tensors. Therefore, all conformal fluids in equilibrium in such boundary geometries are also in perfect equilibrium. In absence of dangerous tensors, we cannot use these boundary geometries to constrain transport coefficients.

Therefore we turn to perfect-Cotton geometries { discussed in Sec. 
\ref{dipax}. }We have found in Sec. 
\ref{hol} that we can uplift these geometries to exact black-hole solutions without naked singularities for generic values of four parameters characterising them. Let us now examine the presence of dangerous tensors in these geometries.

For concreteness, we begin at the third order in derivative expansion. The list of possible dangerous tensors is in \eqref{T3}. We note that $\langle C_{\mu\nu}\rangle$ vanishes in any perfect-Cotton geometry, because the transverse part of $C_{\mu\nu}$ is pure trace, meaning it is proportional to $\Delta_{\mu\nu}$. Therefore, it is not a dangerous tensor in any perfect-Cotton geometry, as a result we cannot constrain the corresponding transport coefficient $\gamma_{(3)1}$. 

We recall from Sec. \ref{perfeq} that we need to evaluate the possible dangerous tensors on-shell, meaning we need to check if they do not vanish when $\mathrm{u} = \xi $. We have shown in App. \ref{Weyl} that in equilibrium, \emph{i.e.} on-shell, the Weyl-covariant derivative $\mathcal{D}_\mu$ reduces to the covariant derivative $\nabla_\mu$. This facilitates our hunt for dangerous tensors.

The first dangerous tensor we encounter is $\langle\mathcal D_\mu W_\nu \rangle$. It is because it is non-vanishing and also it is not conserved, meaning $\nabla^\mu \langle\mathcal D_\mu W_\nu \rangle \neq 0$ in all geometries discussed in Sec. \ref{dipax}. Perfect equilibrium can exist only if the corresponding dangerous transport coefficient $\gamma_{(3)2}$ vanishes. Thus this transport coefficient vanishes for all strongly coupled holographic fluids in the regime of validity of classical gravity approximation.

We can similarly show that infinite number of tensors of the form of $(C^{\alpha\beta}C_{\alpha\beta})^\ell \langle \mathcal D_\mu W_\nu \rangle$, $(V^\alpha V_\alpha)^m \langle \mathcal D_\mu W_\nu \rangle $ and $(W^\alpha W_\alpha)^n \langle \mathcal D_\mu W_\nu \rangle$ for $\ell, m$ and $n$ being arbitrary positive integers, are dangerous tensors in geometries of Sec. \ref{dipax}. We conclude that the infinitely many non-dissipative transport coefficients corresponding to these dangerous tensors should vanish.

At the fourth order in the derivative expansion, we get new kind of dangerous tensors of the form $\langle V_\mu V_\nu \rangle$, $\langle W_\mu W_\nu\rangle$ and $\langle \mathcal D_\mu \mathcal D_\nu (\omega^{\alpha\beta}\omega_{\alpha\beta})\rangle$ in geometries of Sec. \ref{dipax}. This further implies existence of infinite number of dangerous tensors, of the form of $(C^{\alpha\beta}C_{\alpha\beta})^\ell \langle V_\mu V_\nu \rangle$, $(V^\alpha V_\alpha)^m \langle V_\mu V_\nu \rangle $, $(W^\alpha W_\alpha)^n \langle V_\mu V_\nu \rangle$, etc. in the geometries of Sec. \ref{dipax}. Once again this leads us to conclude that infinite number of new dangerous transport coefficients vanish. 

 {To avoid  further technical developments we will not give the exhaustive list of all possible holographic transport coefficients we can constrain using exact black-hole solutions. This will be part of a future work, where systematic perturbations around perfect geometries will allow to probe the non-vanishing transport coefficients.}

We want to conclude this section by arguing that the constraints on transport coefficients derived here cannot be obtained from partition-function \cite{Banerjee:2012iz} or entropy-current \cite{Bhattacharyya:2012nq} based approaches. On the one hand, these methods are very general and independent of holography. On the other hand, our constraints follow from exact solutions of Einstein's equations. In particular, a certain form of duality between the Cotton--York and energy--momentum tensors at the boundary is crucial for us to find these exact solutions. This duality has no obvious direct interpretation in the dual field theory and no obvious connection with general approaches for constraining hydrodynamic transport coefficients. Unfortunately, the general methods mentioned above have been explicitly worked out up to second order in derivative expansion only. However, the first non-trivial constraint in our approach comes at the third order in the derivative expansion. So presently we cannot give an explicit comparison of our technique with these general approaches. It will be interesting to find explicit examples where holographic constraints on transport coefficients discussed here cannot be obtained using different techniques.  Most likely, our results will provide special constraints on the equilibrium partition function for holographic theories.\footnote{We thank Shiraz Minwalla for helpful discussions on this point.}

\section{Outlook}
\label{outlook}

We end here with a discussion on possible future directions. Perhaps the most outstanding question is the classification of all possible perfect geometries for holographic systems. The difficulty in studying this question is to make a formulation which is independent of any ansatz for the metric  {(like the Papapetrou--Randers ansatz we used here),} which will sum over infinite orders in the derivative expansion. It is difficult to show that only a specific ansatz will exhaust all possibilities. In fact it is not clear whether it is necessary to have an exact solution in the bulk in order to have perfect equilibrium in the boundary. There can be derivative corrections to all orders in the bulk metric which cannot be resummed into any obvious form, though such corrections may vanish for the boundary stress tensor.

Recently an interesting technique has been realised for addressing such questions involving the idea of holographic renormalisation-group flow in the fluid/gravity limit \cite{Kuperstein:2013hqa}. In this approach, a fluid is constructed from the renormalised energy--momentum tensor at any hypersurface in the bulk. For a unique hypersurface foliation -- namely the Fefferman--Graham foliation -- the radial evolution of the transport coefficients and hydrodynamic variables is first order and can be constructed without knowing the bulk spacetime metric explicitly. Once this radial evolution is solved, the bulk metric can be constructed from it for a given boundary geometry.

The advantage of this formulation is that the holographic renormalisation-group flow of transport coefficients and hydrodynamic variables automatically knows about the regularity of the horizon. The renormalisation-group flow terminates at the horizon and there exists a unique solution which corresponds to non-relativistic incompressible Navier--Stokes equation at the horizon. This unique solution determines the values of the transport coefficients of the boundary fluid to all orders in the derivative expansion. It is precisely these values which give solutions with regular horizons. Though it has not been proved, this agreement between the renormalisation-group flow and regularity has been checked explicitly for first and second order transport coefficients.

The relevance of this approach to perfect geometries is as follows. In the special case of perfect equilibrium, we know that the boundary fluid should also flow to a fluid in perfect equilibrium at the horizon. The latter can happen only if the boundary geometry is a perfect geometry,  which will impose appropriate restrictions on the fluid kinematics. The question of classification of perfect boundary geometries is thus well posed using deep connections between renormalisation-group flow and horizon regularity -- independently of any specific ansatz. In this approach we will also be able to know the full class of transport coefficients which should necessarily vanish such that perfect equilibrium can exist both at the boundary and the horizon.  

The second immediate question involves further analysis of the black-hole solutions  with at least one extra spatial isometry discussed here. This is particularly necessary for the particular values of the geometric parameters where there exists no curvature singularities in the bulk for all values of the mass. The question is what restricts the mass from being arbitrarily negative -- is it possibly just the requirement that regions of space--time with closed time-like curves should be hidden by horizons? Or do we need new principles? Also we should construct the global thermodynamics of such geometries in detail and investigate if there is anything unusual.

Our guiding principle in searching perfect fluidity is the mass/nut bulk duality, which is a non-linear relationship emerging \emph{a priori} in Euclidean four-dimensional gravity. Its manifestation in Lorentzian geometries is holographic and operates linearly via the Cotton/energy--momentum duality on the $2+1$-dimensional boundary; it is a kind of duality relating the energy density with the vorticity, when the later is non-trivial \emph{i.e.} when the Cotton--York tensor is non-vanishing. This relationship should be further investigated as it provides another perspective on gravity duality \cite{BunHen}.

Finally, it will be interesting to find exact solutions in the bulk with matter fields corresponding to steady states in the boundary. These steady states will be sustained by non-normalisable modes of the bulk matter fields. Perhaps the simplest and the most interesting possibility is adding axion fields with standard kinetic term in the bulk which couple also to the Gauss--Bonnet term. Such bulk actions have been studied recently \cite{Saremi:2011ab, Delsate:2011qp, Jensen:2011xb, Liu:2012zm}. In fact, it has been shown that this leads to simple mechanism for generating vortices in the boundary spontaneously. These simple vortices describe transitions in the $\theta$ vacuum across an edge and support edge currents. It will be interesting to see if there could be non-trivial exact solutions in the bulk describing more general steady state vortex configurations in the bulk. The relevant question analogous to the one studied in this work will be which boundary geometries and axionic configurations can sustain steady vortex configurations. 
\section*{Acknowledgements}
{\small The authors wish to thank   G. Barnich, M. Caldarelli, {J. Gath}, S. Katmadas, D. Klemm, R.G. Leigh, S.~Minwalla, N. Obers, K. Sfetsos  and Ph. Spindel  for a number of {interesting} discussions. We also thank M. Caldarelli and K. Jensen for very useful comments on the first version of the manuscript. P.M.P., K.S. and A.C.P. would like to thank each others home institutions for hospitality, where part of this work was developed. 
In addition A.C.P.  and K.S. thank the Laboratoire de Physique Th\'eorique  of the Ecole Normale Sup\'erieure for hospitality. The present work was completed during the 2013 Corfu EISA Summer Institute.
This research was supported by the LABEX P2IO, the ANR contract  05-BLAN-NT09-573739, the ERC Advanced Grant  226371 and the ITN programme PITN-GA-2009-237920. The work
of A.C.P. was partially supported by the Greek government research program \textsl{AdS/CMT -- Holography and
Condensed Matter Physics} (ERC -- 05), MIS 37407. The work of K.S. has been supported by  \textsl{Actions de recherche concert\'ees (ARC)} de la 
\textsl{Direction g\'en\'erale
de l'Enseignement non obligatoire et de la Recherche scientifique Direction de la Recherche scientifique Communaut\'e
fran\c{c}aise de Belgique}, and by IISN-Belgium (convention
4.4511.06). }

\appendix

\section{On vector-field congruences}\label{vfc}

We consider a manifold endowed with a space--time metric of the generic form
\begin{equation}\label{Dmet}
\mathrm{d}s^2 =g_{\mu\nu}\mathrm{d}\mathrm{x}^\mu \mathrm{d}\mathrm{x}^\nu= \eta_{ab}\mathrm{e}^a \mathrm{e}^b.
\end{equation}
We will use $a,b,c,\ldots =0,1,\ldots, D-1$ for transverse Lorentz indices along with $\alpha,\beta,\gamma=1,\ldots, D-1$. 
Coordinate indices will be denoted $\mu,\nu,\rho, \ldots$ for space--time $\mathrm{x}\equiv (t,x)$ and  $i,j,k, \ldots$ for 
spatial $x$ directions. 
Consider now an arbitrary time-like vector field $\mathrm{u}$, normalised as  $u^\mu u_\mu=-1$,  later identified with the fluid velocity. Its integral curves define a congruence which is characterised by its acceleration, shear, expansion and vorticity (see e.g. \cite{Ehlers:1993gf,vanElst:1996dr}):
\begin{equation}
\label{def1}
\nabla_{\mu} u_\nu=-u_\mu a_\nu +\frac{1}{D-1}\Theta \Delta_{\mu\nu}+\sigma_{\mu\nu} +\omega_{\mu\nu}
\end{equation}
with\footnote{Our conventions are: $A_{(\mu\nu)}=\nicefrac{1}{2}\left(A_{\mu\nu}+A_{\nu\mu}\right)$ and $A_{[\mu\nu]}=\nicefrac{1}{2}\left(A_{\mu\nu}-A_{\nu\mu}\right)$.}
\begin{eqnarray}
a_\mu&=&u^\nu\nabla_\nu u_\mu, \quad
\Theta=\nabla_\mu u^\mu, \label{def21}\\
\sigma_{\mu\nu }&=&\frac{1}{2} \Delta_\mu^{\hphantom{\mu}\rho } \Delta_\nu ^{\hphantom{\nu }\sigma}\left(
\nabla_\rho  u_\sigma +\nabla_\sigma  u_\rho 
\right)-\frac{1}{D-1} \Delta_{\mu\nu }\Delta^{\rho \sigma } \nabla_\rho  u_\sigma  \label{def22}\\
&=& \nabla_{(\mu} u_{\nu )} + a_{(\mu} u_{\nu )} -\frac{1}{D-1} \Delta_{\mu\nu } \nabla_\rho  u^\rho  ,
\label{def23}\\
\omega_{\mu\nu }&=&\frac{1}{2} \Delta_\mu^{\hphantom{\mu}\rho } \Delta_\nu ^{\hphantom{\nu }\sigma }\left(
\nabla_\rho  u_\sigma -\nabla_\sigma  u_\rho 
\right)= \nabla_{[\mu} u_{\nu ]} + u_{[\mu} a_{\nu ]}.\label{def24}
\end{eqnarray}
The latter allows to define the vorticity form as
\begin{equation}\label{def3}
2\omega=\omega_{\mu\nu }\, \mathrm{d}\mathrm{x}^\mu\wedge\mathrm{d}\mathrm{x}^\nu  =\mathrm{d}\mathrm{u} +
\mathrm{u} \wedge\mathrm{a}\, .
\end{equation}
The time-like vector field $\mathrm{u}$  has been used to decompose any tensor field on the manifold in transverse and longitudinal components with respect to itself.  The decomposition is performed by introducing the longitudinal and transverse projectors:
\begin{equation}
\label{proj}
U^\mu_{\hphantom{\mu}\nu } = - u^\mu u_\nu , \quad \Delta^\mu_{\hphantom{\mu}\nu } =  u^\mu u_\nu  + \delta^\mu_{\nu },
\end{equation}
where $\Delta_{\mu \nu }$ is also the induced metric on the surface orthogonal  to $\mathrm{u}$. The projectors satisfy the usual identities:
\begin{equation}
U^\mu_{\hphantom{\mu}\rho } U^\rho _{\hphantom{\rho }\nu } = U^\mu_{\hphantom{\mu}\nu },\quad U^\mu_{\hphantom{\mu}\rho } \Delta^\rho _{\hphantom{\rho }\nu }  =   0 , \quad \Delta^\mu_{\hphantom{\mu}\rho } \Delta^\rho _{\hphantom{\rho }\nu }  =   \Delta^\mu_{\hphantom{\mu}\nu } ,\quad U^\mu_{\hphantom{\mu}\mu}=1, \quad \Delta^\mu_{\hphantom{\mu}\mu}=D-1,
\end{equation}
and similarly:
\begin{equation}
u^\mu a_\mu=0, \quad u^\mu \sigma_{\mu\nu }=0,\quad u^\mu \omega_{\mu\nu }=0, \quad u^\mu \nabla_\nu  u_\mu=0, \quad \Delta^\rho _{\hphantom{\rho }\mu} \nabla_\nu  u_\rho  =\nabla_\nu  u_\mu.
\end{equation}

\section{Weyl-covariant traceless transverse tensors in hydrodynamics}\label{Weyl}

The presentation here will mostly follow \cite{Rangamani:2009xk}.
It is possible to express the hydrodynamics tensors in a manifest Weyl-covariant way. To do so, we first need to define a torsionless Weyl-connection $\nabla_\rho ^{\mathrm{Weyl} }$ over $\left( \mathcal M, \mathcal C \right)$, where $\mathcal M$ is the three-dimensional manifold and $\mathcal C$ is the conformal class of metrics on the manifold:
\begin{equation}
\label{WeylConnection}
\nabla_\rho ^{\mathrm{Weyl} } g_{\mu \nu} = 2 \mathcal A_\rho g_{\mu \nu}. 
\end{equation}
In the latter, $g_{\mu \nu}$ is any metric in the conformal class $\mathcal C$ and $\mathcal A_\mu$ is a one-form. Using the Weyl-connection it is possible to define a Weyl-covariant derivative $\mathcal D_\mu ^{\mathrm{Weyl}} = \nabla _\mu + \omega \mathcal A_\mu$, where $\omega$ is the conformal weight of the tensor on which the derivative is acting. If the behavior of a tensor $\mathcal Q_{\nu \ldots} ^{\mu \ldots}$ under conformal transformation is $\mathcal Q_{\nu \ldots} ^{\mu \ldots} = 
e^{- \omega \phi} \tilde {\mathcal Q} _{\nu \ldots} ^{\mu \ldots}$, then under the same transformation the derivative will transform in a covariant way, that is $\mathcal D_\rho ^{\mathrm{Weyl} } \mathcal Q_{\nu \ldots} ^{\mu \ldots} = e^{- \omega \phi} \mathcal D_\rho ^{\mathrm{Weyl} } \tilde {\mathcal Q} _{\nu \ldots} ^{\mu \ldots}$. The explicit expression of the Weyl-covariant derivative is given by
\begin{equation}
\begin{split}
\mathcal D_\rho \mathcal Q_{\nu \ldots} ^{\mu \ldots}  &\equiv \nabla_\rho \mathcal Q_{\nu \ldots} ^{\mu \ldots}  + \omega \mathcal A_\rho \mathcal Q_{\nu \ldots} ^{\mu \ldots}  \\
&+ \left( g_{\rho \sigma} \mathcal A^\mu - \delta_\rho ^\mu \mathcal A_\sigma - \delta_\sigma ^\mu \mathcal A_\rho \right) \mathcal Q_{\nu\ldots} ^{\sigma \ldots} + \cdots
\\
&- \left( g_{\rho \nu} \mathcal A^\sigma - \delta_\rho ^\sigma \mathcal A_\nu - \delta^\sigma _\nu \mathcal A_\rho \right) \mathcal Q^{\mu \ldots} _{\sigma \ldots}  + \cdots
\end{split}
\end{equation}
From \eqref{WeylConnection} it follows immediately that the Weyl-covariant derivative is metric-compatible:
\begin{equation}
\mathcal D_\rho g_{\mu \nu}=0,
\end{equation}
since the metric tensor has weight $\omega = -2$. The connection one-form $\mathcal A_\mu$ is uniquely determined by demanding the Weyl-covariant derivative of the velocity of the fluid to be transverse and traceless
\begin{equation} 
u^\rho \mathcal D_\rho u^\nu = 0, \quad \mathcal D_\rho u^\rho =0, 
\end{equation}
which imply
\begin{equation}
A_\mu = u^\rho \nabla_\rho u_\mu - \frac{1}{D-1} u_\mu \nabla ^\rho u_\rho \equiv a_\mu - \frac{1}{D-1} \Theta \, u_\mu. 
\end{equation}
From the latter it is straightforward to see that for all the configurations we considered $\mathcal A_\mu=0$, since both the acceleration and the expansion rate are vanishing, and thus the Weyl-covariant derivative reduces to the normal derivative.

\section{Explicit bulk solutions}\label{solutions}

The dual of perfect-Cotton boundary geometries can be written as an exact solution of Einstein's equations. Such solutions are different depending on the value of $c_4$ and on the geometry of the horizon. For non-vanishing $c_4$, all these metrics belong to the Pleba\~{n}ski--Demia\`{n}ski type $\text{D}$ class \cite{PD}, without acceleration
and/or rotation parameters.\footnote{For non-vanishing acceleration in the presence of rotation, the Pleba\~{n}ski--Demia\`{n}ski boundary \emph{is not} perfect-Cotton. The holographic fluid properties of this exact stationary Einstein metric deserve further investigation.}

\boldmath
\subsection*{Non-vanishing $c_4$: Kerr--Taub--NUT metrics}
\unboldmath

We start from the boundary metrics studied in Sec. \ref{dipax} and uplift them using \eqref{bulkmetricBG}. 
\paragraph*{Spherical ($\nu=1$)}    
We set
\begin{eqnarray}
c_1 &=& 2  (a-n), \nonumber\\
c_2 &=& 2 a  (-1 + a^2  - 4 a n  ), \nonumber\\
c_3 &=&  -1 + 5 a^2  - 12 an.
\end{eqnarray}
By doing this, we recover the spherical-horizon Kerr--Taub--NUT metric \cite{Chen:2006xh}:
\begin{equation}
\mathrm d s^2 = \frac{\rho^2}{\Delta_r}\mathrm{d}r^2 
- \frac{\Delta_r}{\rho^2}\left(\mathrm{d}t + \beta\mathrm{d}{ \varphi} \right)^2
+\frac{\rho^2}{\Delta_\vartheta}\mathrm{d}\vartheta^2 +  \frac{\sin^2\vartheta\Delta_\vartheta}{\rho^2}\left(a \mathrm{d}t +  \alpha \mathrm{d}{\varphi} \right)^2,
\end{equation}
with
\begin{eqnarray}
\rho^2 &=& r^2  + (n- a \cos\vartheta)^2, \\
\Delta_r &=&  r^4 +r^2 (1 + a^2  + 6 n^2) -2 M r + (a^2 - n^2)  (1 + 3 n^2 ) , \\
\Delta_\vartheta &=&  1 + a \cos\vartheta(4n - a \cos\vartheta), \\
\beta &=& -b(\theta)  = -\frac{2 (a - 2 n + a \cos\vartheta)}{\Xi}\sin^2\nicefrac{\vartheta}{2},\\
\alpha &=& -\frac{r^2+(n-a)^2}{\Xi}, \\
\Xi &=& 1- a^2 .
 \end{eqnarray}

\paragraph*{Flat ($\nu=0$)}
We set
\begin{eqnarray}
c_1 &=& 2  (a-n), \nonumber\\
c_2 &=& 2 a^2  (a -4n), \nonumber\\
c_3 &=& a (5 a - 12 n).
\end{eqnarray}
and get the flat-horizon Kerr--Taub--NUT metric:
\begin{equation}
\label{fh-KNT}
\mathrm d s^2 = \frac{\rho^2}{\Delta_r}\mathrm{d}r^2 
- \frac{\Delta_r}{\rho^2}\left(\mathrm{d}t + \beta\mathrm{d}\varphi \right)^2
+\frac{\rho^2}{\Delta_\sigma}\mathrm{d}\sigma^2 +  \frac{\sigma^2\Delta_\sigma}{\rho^2}\left(a^2 (a- 4n) \mathrm{d}t +  \alpha \mathrm{d}\varphi \right)^2,
\end{equation}
with
\begin{eqnarray}
\rho^2 &=& r^2 + \frac{1}{4} \left(2 a - 2 n + a^2  \sigma^2 (a- 4n)\right)^2, \\
\Delta_r &=&  r^4 + r^2 (a^2 + 6 n^2)  - 2Mr + 3 n^2 (a^2 - n^2)  , \\
\Delta_\sigma &=& \frac{(2 + a^2  \sigma^2) (8 - 24 a n  \sigma^2 + 
   a^4  \sigma^4 - 8 a^3 n  \sigma^4 + 
   2 a^2  \sigma^2 (3 + 8 n^2  \sigma^2))}{16}, \\
\beta &=& -b(\theta) \,\, \, = \,\,\, \frac{\sigma^2}{4} \left(4 (n- a) + a^2  \sigma^2 (4n -a)\right) ,\\
\alpha &=& r^2+(n-a)^2 .
 \end{eqnarray}
It seems that this metric was never quoted in the literature. It provides the AdS generalisation of the asymptotically flat metric of \cite{demianski:1966}.

\paragraph*{Hyperbolic ($\nu=-1$)}
We set
\begin{eqnarray}
c_1 &=& 2  (a-n), \nonumber\\
c_2 &=& 2 a  (1 + a^2  - 4 a n  ), \nonumber\\
c_3 &=&  1 + 5 a^2  - 12 an.
\end{eqnarray}
and obtain the hyperbolic-horizon Kerr--Taub--NUT metric (also mentioned in \cite{Ortin}):
\begin{equation}
\label{KTNhyp}
\mathrm d s^2 = \frac{\rho^2}{\Delta_r}\mathrm{d}r^2 
- \frac{\Delta_r}{\rho^2}\left(\mathrm{d}t + \beta\mathrm{d}{ \varphi} \right)^2
+\frac{\rho^2}{\Delta_\theta}\mathrm{d}\theta^2 +  \frac{\sinh^2\theta\Delta_\theta}{\rho^2}\left(a  \mathrm{d}t +  \alpha \mathrm{d}{ \varphi} \right)^2,
\end{equation}
with
\begin{eqnarray}
\rho^2 &=& r^2  + (n- a \cosh\theta)^2, \\
\Delta_r &=&  r^4 + 
 r^2  (-1 + a^2  + 6 n^2 )-2 M r + (a^2 - n^2)  (-1 + 3 n^2), \\
\Delta_\theta &=&  1 - a \cosh\theta(4n - a \cosh\theta), \\
\beta &=& -b(\theta) = -\frac{2 (a - 2 n + a \cosh\theta)}{Z}\sinh^2\nicefrac{\theta}{2},\\
\alpha &=& \frac{r^2+(n-a)^2}{Z}, \\
Z &=& 1+ a^2.
 \end{eqnarray}
 
 \boldmath
 \subsection*{Vanishing $c_4$}
 \unboldmath
When $c_4 = 0$, the bulk metric is obtained in Boyer--{Lindqvist} form from \eqref{bulkmetric1} by doing the following coordinate transformations
\begin{eqnarray}
\mathrm{d}\tilde{t} &=& \mathrm{d}t -\frac{4(c_1^2 + 4 r^2) }{3c_1^4  - 4c_1^2 (c_3 + 6r^2)-16 (c_2 ^2 c_5 -2M r - c_3 r^2 + r^4)}\mathrm{d}r, \\
\mathrm{d}\tilde{y} &=& \mathrm{d}y + \frac{16 c_2}{3c_1^4  - 4c_1^2 (c_3 + 6r^2)-16 (c_2 ^2 c_5 -2M r - c_3 r^2 + r^4)}\mathrm{d}r.
\end{eqnarray}
The metric is explicitly given by (we replace $\tilde{r}$ and $\tilde{y}$ with $r$ and $y$):
\begin{equation}\label{bulkmetricBGa}
\mathrm{d}s^2 = \frac{\rho^2}{\Delta_r}\mathrm{d}r^2 
- \frac{\Delta_r}{\rho^2}\left(\mathrm{d}t + \beta\mathrm{d}y \right)^2
+\frac{\rho^2}{\Delta_x}\mathrm{d}x^2 +  \frac{\Delta_x}{\rho^2}\left(c_2 \mathrm{d}t - \alpha \mathrm{d}y \right)^2,
\end{equation}
where
\begin{eqnarray}
\rho^2 &=& r^2 + \frac{q^2}{4} \, =\, r^2 + \frac{(c_1 + 2 c_2x)^2}{4}, \\
\Delta_r &=& - \frac{1}{16}\left(3c_1^4  - 4c_1^2 (c_3 + 6r^2)-16 (c_2 ^2 c_5 -2M r - c_3 r^2 + r^4)\right), \\
\Delta_x &=& G  = c_5 + c_3 x^2 + 2 c_1 c_2 x^3 + c_2^2 x^4, \\
\alpha &=& -\frac{1}{4} \left(c_1^2 + 4 r^2\right), \\
\beta &=& -b  =  - c_1x - c_2 x^2.
\end{eqnarray}
According to our knowledge, solutions of these kind are not known in literature except for the special case where $c_1$, $c_2$, $c_3$ and $c_5$ are given by
 \begin{equation}
 \label{ortin}
c_1 = 2 n, \quad
c_2 = a , \quad c_3 =  0\quad \text{and}\quad
c_5 =  1,
\end{equation}
with $n$ being the nut charge and $a$ being the angular momentum. In this case we recover the flat-horizon solution of \cite{Ortin}, which is however different from the above flat-horizon solution \eqref{fh-KNT}, or, at $n=0$, the rotating topological black hole of \cite{Klemm:1997ea}. 

 \subsection*{Geometries with conformally flat boundarties}

As discussed at the end of Secs. \ref{nvanc4} and \ref{dipvanc4}, the boundary geometries become conformally flat for some specific values of the parameters: for non-vanishing $c_4$ this happens when \eqref{cvan} is satisfied, whereas for $c_4=0$ the requirement is \eqref{cvanc4van}.

Whenever the nut charge vanishes ($n=0$ or $c_1=0$), the boundary is conformally flat and the bulk geometries correspond to standard rotating black holes. However, there are more interesting situations, which we will not analyse extensively here. We simply quote the one reached with $c_4\neq 0$, $\nu = -1, a=0, n=-\nicefrac{1}{2}$, the boundary of which is $\mathrm{AdS}_3$. The bulk (given in \eqref{KTNhyp}) is actually a rotating hyperboloid black brane. This geometry can also be obtained by zooming around the north pole of the ultraspinning (\emph{i.e.} $a=1$) spherical-horizon ($\nu=1, n=0$) black hole (see \cite{NewPaper, Caldarelli:2008pz}).



\begin{thebibliography}{99}

\bibitem{Romatschke:2009im}
  P.~Romatschke,
  Int.\ J.\ Mod.\ Phys.\ \textbf{E19} (2010) 1
  [arXiv:0902.3663 [hep-ph]].


\bibitem{Kovtun:2012rj}
  P.~Kovtun,
  J.\ Phys.\  {\bf A45} (2012) 473001
  [arXiv:1205.5040 [hep-th]].
  
\bibitem{Hubeny}
  V.E.~Hubeny, S.~Minwalla and M.~Rangamani,
  [arXiv:1107.5780 [hep-th]].

\bibitem{Rangamani:2009xk} 
  M.~Rangamani,
  Class.\ Quant.\ Grav.\  {\bf 26}, 224003 (2009)
  [arXiv:0905.4352 [hep-th]].
 
\bibitem{Leigh:2007wf} 
  R.G.~Leigh and A.C.~Petkou,
  JHEP {\bf 0711}, 079 (2007)
  [arXiv:0704.0531 [hep-th]].

\bibitem{Mansi:2008br} 
  D.S.~Mansi, A.C.~Petkou and G.~Tagliabue,
  Class.\ Quant.\ Grav.\  {\bf 26} (2009) 045008
  [arXiv:0808.1212 [hep-th]].
  
\bibitem{Mansi:2008bs} 
  D.S.~Mansi, A.C.~Petkou and G.~Tagliabue,
  Class.\ Quant.\ Grav.\  {\bf 26} (2009) 045009 
  [arXiv:0808.1213 [hep-th]].

\bibitem{deHaro:2008gp}
  S.~de Haro,
  JHEP {\bf 0901} (2009) 042
  [arXiv:0808.2054 [hep-th]].


\bibitem{Miskovic:2009bm}
  O.~Miskovic and R.~Olea,
  Phys.\ Rev.\ {\bf D79} (2009) 124020
  [arXiv:0902.2082 [hep-th]].


\bibitem{PD} 
J.F. Pleba\~{n}ski and M. Demia\`{n}ski, Ann. Phys. (NY) \textbf{98} (1976) 98.

\bibitem{Banerjee:2012iz} 
  N.~Banerjee, J.~Bhattacharya, S.~Bhattacharyya, S.~Jain, S.~Minwalla and T.~Sharma,
  JHEP {\bf 1209} (2012) 046 
  [arXiv:1203.3544 [hep-th]].

\bibitem{Jensen:2012jh} 
  K.~Jensen, M.~Kaminski, P.~Kovtun, R.~Meyer, A.~Ritz and A.~Yarom,
  Phys.\ Rev.\ Lett.\  {\bf 109} (2012) 101601
  [arXiv:1203.3556 [hep-th]].

\bibitem{Bhattacharyya:2012nq} 
  S.~Bhattacharyya,
  JHEP {\bf 1207} (2012) 104
  [arXiv:1201.4654 [hep-th]].
  
  
\bibitem{Bhattacharyya:2007vs} 
  S.~Bhattacharyya, S.~Lahiri, R.~Loganayagam and S.~Minwalla,
  JHEP {\bf 0809} (2008)  054
  [arXiv:0708.1770 [hep-th]].

\bibitem{Moore:2010bu} 
  G.D.~Moore and K.A.~Sohrabi,
  Phys.\ Rev.\ Lett.\  {\bf 106} (2011)  122302
  [arXiv:1007.5333 [hep-ph]].

\bibitem{Loga} 
  R.~Loganayagam,
  JHEP {\bf 0805} (2008) 087
  [arXiv:0801.3701 [hep-th]].

\bibitem{Papapetrou}
A. Papapetrou, 
Ann. Inst. H. Poincar\'e {\bf A4} (1966) 83.

\bibitem{Randers}
G. Randers, 
Phys. Rev. {\bf 59}
(1941) 195.

\bibitem{Gibbons}
  G.W.~Gibbons, C.A.R.~Herdeiro, C.M.~Warnick and M.C.~Werner,
  Phys.\ Rev.\ {\bf D79} (2009) 044022
  [arXiv:0811.2877 [gr-qc]].

\bibitem{Leigh:2011au}
  R.G.~Leigh, A.C.~Petkou and P.M.~Petropoulos,
  Phys.\ Rev. {\bf D85} (2012) 086010
  [arXiv:1108.1393 [hep-th]].
  
\bibitem{LPP2}
  R.G.~Leigh, A.C.~Petkou and P.M.~Petropoulos,
  JHEP \textbf{1211} (2012) 121
   [arXiv:1205.6140 [hep-th]].

\bibitem{NewPaper}
M.M. Caldarelli, R.G. Leigh, A.C. Petkou, P.M. Petropoulos, V. Pozzoli and K. Siampos, 
Proc. of Science \textbf{Corfu11} (2012) 076  [arXiv:1206.4351 [hep-th]].

\bibitem{Moutsopoulos:2011ez}
  G.~Moutsopoulos and P.~Ritter,
  Gen.\ Rel.\ Grav.\  {\bf 43} (2011) 3047
  [arXiv:1103.0152 [hep-th]].
  
\bibitem{Chow:2009km}
  D.D.K.~Chow, C.N.~Pope and E.~Sezgin,
  Class.\ Quant.\ Grav.\  {\bf 27} (2010) 105001
  [arXiv:0906.3559 [hep-th]].
  

    \bibitem{rayPRD80}
  A.K. Raychaudhuri and S.N. Guha Thakurta,
  Phys. Rev. \textbf{D22} (1980) 802.
   
      \bibitem{rebPRD83}
M.J. Rebou\c{c}as and J. Tiomno,
  Phys. Rev. \textbf{D28} (1983) 1251.
  
  \bibitem{SR68}
  M.M. Som and  A.K. Raychaudhuri, 
  Proc. R. Soc. London \textbf{A304} (1968) 81.

\bibitem{cs}
S. Deser, R. Jackiw and S. Templeton, 
Ann. Phys. \textbf{140} (1982) 372; Erratum-ibid. \textbf{185}
(1988) 406; 
Phys. Rev. Lett. \textbf{48} (1982) 975.


\bibitem{Anninos:2008fx}
  D.~Anninos, W.~Li, M.~Padi, W.~Song and A.~Strominger,
  JHEP {\bf 0903} (2009) 130
  [arXiv:0807.3040 [hep-th]].

\bibitem{Anninos:2011vd}
  D.~Anninos, S.~de Buyl and S.~Detournay,
  JHEP {\bf 1105} (2011) 003
  [arXiv:1102.3178 [hep-th]].

   \bibitem{Grumiller}  
 D.~Grumiller and W.~Kummer,
  Annals Phys.\  {\bf 308} (2003) 211
  [hep-th/0306036].

 \bibitem{Guralnik}  
   G.~Guralnik, A.~Iorio, R.~Jackiw and S.Y.Pi,
  Annals Phys.\  {\bf 308} (2003) 222
  [hep-th/0305117].


\bibitem{Hawking:1998kw}
  S.W.~Hawking, C.J.~Hunter and M.~Taylor,
  Phys.\ Rev.\  {\bf D59} (1999) 064005
  [hep-th/9811056].

\bibitem{Ortin}
  N.~Alonso-Alberca, P.~Meessen and T.~Ortin,
  Class.\ Quant.\ Grav.\  {\bf 17} (2000) 2783
  [arXiv:0003071 [hep-th]].


\bibitem{SS}
K.~Skenderis and S.N.~Solodukhin,
  Phys.\ Lett.\  {\bf B472} (2000) 316
  [hep-th/9910023].
  
   \bibitem{GP} J.B.  Griffiths and J. Podolsk\'y, \textsl{Exact space--times in Einstein's general relativity}, Cambridge University Press, 2009.

\bibitem{Eling:2013sna}
  C.~Eling and Y.~Oz,
  {JHEP {\bf 1311} (2013) 079}
  [arXiv:1308.1651 [hep-th]].

 \bibitem{FG}
  C. Fefferman and C.R. Graham,
    arXiv:0710.0919 [math.DG].


 \bibitem{LeBrun82}
C.R. Lebrun, 
Proc. R. Soc. Lond. \textbf{A380} (1982) 171.

\bibitem{Pedepoon90}
H. Pedersen and Y.S. Poon, 
Class.\ Quant.\ Grav.\  {\bf 7} (1990) 1707.

\bibitem{Tod90}
K.P. Tod, 
Class.\ Quant.\ Grav.\  {\bf 8} (1991) 1049.

\bibitem{Tod94}
K.P. Tod, 
Phys. Lett. \textbf{A190} (1994) 221.

\bibitem{Hitchin95}
N.J. Hitchin, 
J. Diff. Geom. \textbf{42} (1995) 30.

\bibitem{Bhattacharyya:20082}
  S.~Bhattacharyya, R.~Loganayagam, I. Mandal, S.~Minwalla and A. Sharma,
  JHEP {\bf 0812} (2008) 116
  [arXiv:0809.4272 [hep-th]].


\bibitem{Bhattacharyya:2008ji}
  S.~Bhattacharyya, R.~Loganayagam, S.~Minwalla, S.~Nampuri, S.P.~Trivedi and S.R.~Wadia,
  JHEP {\bf 0902} (2009) 018
  [arXiv:0806.0006 [hep-th]].
  
  \bibitem{Clarkson:2002uj}
  R.~Clarkson, L.~Fatibene and R.B.~Mann,
  Nucl.\ Phys.\ {\bf B652} (2003) 348
  [hep-th/0210280].

\bibitem{Kuperstein:2013hqa} 
 S.~Kuperstein and A.~Mukhopadhyay,
  {JHEP {\bf 1311} (2013) 086}
  [arXiv:1307.1367 [hep-th]].

\bibitem{BunHen}
 C.~Bunster, M.~Henneaux and S.~Hortner,
 arXiv:1301.5496 [hep-th].

\bibitem{Saremi:2011ab} 
  O.~Saremi and D.T.~Son,
  JHEP {\bf 1204}, 091 (2012)
  [arXiv:1103.4851 [hep-th]].
  
  \bibitem{Delsate:2011qp}
  T.~Delsate, V.~Cardoso and P.~Pani,
  JHEP {\bf 1106} (2011) 055
  [arXiv:1103.5756 [hep-th]].
  

\bibitem{Jensen:2011xb} 
  K.~Jensen, M.~Kaminski, P.~Kovtun, R.~Meyer, A.~Ritz and A.~Yarom,
  JHEP {\bf 1205} (2012) 102 
  [arXiv:1112.4498 [hep-th]].

\bibitem{Liu:2012zm} 
  H.~Liu, H.~Ooguri, B.~Stoica and N.~Yunes,
  Phys.\ Rev.\ Lett.\  {\bf 110} (2013) 211601
  [arXiv:1212.3666 [hep-th]].
 

\bibitem{Ehlers:1993gf}
  J.~Ehlers,
  Gen.\ Rel.\ Grav.\  {\bf 25} (1993) 1225.

\bibitem{vanElst:1996dr}
  H.~van Elst and C.~Uggla,
  Class.\ Quant.\ Grav.\  {\bf 14} (1997) 2673
  [gr-qc/9603026].


  
\bibitem{Chen:2006xh}
  W.Chen, H.~Lu and C.N.~Pope,
  Class.\ Quant.\ Grav.\  {\bf 23} (2006) 5323
  [hep-th/0604125].
  
  
  \bibitem{demianski:1966}
M. Demia\`{n}ski and E.T. Newman, 
Bulletin de l'Acad\'emie Polonaise des Sciences, \textbf{XIV} (1966) 653.
 
 
\bibitem{Caldarelli:2008pz}
  M.M.~Caldarelli, R.~Emparan and M.J.~Rodriguez,
  JHEP {\bf 0811} (2008) 011
  [arXiv:0806.1954 [hep-th]].

\bibitem{Klemm:1997ea}
  D.~Klemm, V.~Moretti and L.~Vanzo,
  Phys.\ Rev.\ {\bf D57} (1998) 6127
   [Erratum-ibid.\ {\bf D60} (1999) 109902]
  [gr-qc/9710123].

\end{thebibliography}
\end{document}